\documentclass[twocolumn,floatfix,longbibliography]{revtex4}
\usepackage[utf8]{inputenc}
\usepackage{amsfonts,amsmath,amssymb,amsthm,graphicx,epstopdf,verbatim,physics,dsfont}
\usepackage{placeins}
\usepackage{rotating}
\usepackage{epstopdf}
\usepackage[dvipsnames]{xcolor}
\usepackage[normalem]{ulem}
\usepackage{hyperref}

\DeclareMathOperator{\delop}{\hat{\delta}}

\DeclareMathOperator{\Oop}{\hat{O}}
\DeclareMathOperator{\Ham}{\hat{H}}
\DeclareMathOperator{\Rho}{\hat{\rho}}
\DeclareMathOperator{\Gamop}{\hat{\Gamma}}
\newcommand{\jump}{\xrightarrow{\rm jump}}

\DeclareMathOperator{\sop}{\hat{\sigma}}

\newcommand{\ess}[4]{\ev{\sop_{#1}^{#2}\sop_{#3}^{#4}}}
\newcommand{\es}[2]{\ev{\sop_{#1}^{#2}}}
\newcommand{\eff}[4]{\ev{\delop_{#1}^{#2}\delop_{#3}^{#4}}}
\newcommand{\dZ}[1]{\dd{Z}_{#1}}

\graphicspath{{FinalFigs}}

\begin{document}


\title{Quantum and classical correlations in open quantum-spin lattices via truncated-cumulant trajectories}

\author{Wouter Verstraelen}
\thanks{These authors contributed equally}
\affiliation{Division of Physics and Applied Physics, School of Physical and Mathematical Sciences,
Nanyang Technological University, Singapore 637371, Singapore}

\affiliation{TQC, Universiteit Antwerpen, Universiteitsplein 1, B-2610 Antwerpen, Belgium}

\author{Dolf Huybrechts}
\thanks{These authors contributed equally}
\affiliation{Univ Lyon, Ens de Lyon, CNRS, Laboratoire de Physique, F-69342 Lyon, France}
\affiliation{TQC, Universiteit Antwerpen, Universiteitsplein 1, B-2610 Antwerpen, Belgium}

\author{Tommaso Roscilde}
\affiliation{Univ Lyon, Ens de Lyon, CNRS, Laboratoire de Physique, F-69342 Lyon, France}

\author{Michiel Wouters}
\affiliation{TQC, Universiteit Antwerpen, Universiteitsplein 1, B-2610 Antwerpen, Belgium}

\begin{abstract}

The study of quantum many-body physics in Liouvillian open quantum systems becomes increasingly important with the recent progress in experimental control on dissipative systems and their exploitation for technological purposes. A central question in open quantum systems concerns the fate of quantum correlations, and the possibility of controlling them by engineering the competition between the Hamiltonian dynamics and the coupling of the system to a bath. Such a question is very challenging from a theoretical point of view, as numerical methods faithfully accounting for quantum correlations are either relying on exact diagonalization, limiting drastically the sizes that can be treated numerically; or on approximations on the range or strength of quantum correlations, associated to the choice of a specific Ansatz for the density matrix. In this work we propose a new method to treat open quantum-spin lattices, based on stochastic quantum trajectories for the solution of the open-system dynamics. Along each trajectory, the hierarchy of equations of motion for many-point spin-spin correlators is truncated to a given finite order,  assuming that  multivariate $k$-th order cumulants vanish for $k$ exceeding a cutoff $k_c$. This scheme allows one to track the evolution of quantum spin-spin correlations up to order $k_c$ for all length scales. We validate this approach in the paradigmatic case of the dissipative phase transitions of the 2D XYZ lattice subject to spontaneous decay. We convincingly assess the existence of steady-state phase transitions from paramagnetic to ferromagnetic, and back to paramagnetic, upon increasing one of the Hamiltonian spin-spin couplings; as well as the classical Ising nature of such transitions. Moreover, the approach allows us to show the presence of significant quantum correlations in the vicinity of the dissipative critical point, and to unveil the presence of spin squeezing, which can be proven to be a tight lower bound to the quantum Fisher information. 
\end{abstract}

\date{\today}

\maketitle

\section{Introduction}


After many decades of remarkable successes in describing nature at the microscopic level, and in providing foundational contributions to modern technology, the frontier of research on quantum mechanics has turned in the last decade to the hitherto impossible creation and manipulation of quantum entanglement at large scales, constituting the basis of the so-called second quantum revolution \cite{Dowling2003,Preskill2012Arxiv}. The main technological rewards of this revolution are expected to lie in sensing, communication and information processing \cite{acinetal_roadmap_2018,Altmanetal2021}. A most important challenge in this endeavor stems from the interaction of quantum systems with their environment, which generally destroys quantum superpositions, namely the fundamental resource for quantum technology tasks. This calls for a detailed understanding of the interaction of complex quantum systems with their environment; and for the devising of strategies to overcome decoherence. The most obvious remedy is to limit the interaction with the environment as much as possible; but an alternative strategy is to investigate the possibility of creating interesting quantum states by engineering the environment itself \cite{Plenio99,VerstraeteNATPH2009}.

A promising setting for the non-unitary manipulation of quantum states is that of a phase transition in a driven-dissipative many-body system \cite{MingantiPRA18_Spectral,Sieberer_2016,VerstraelenPRR20,KrimerPRL19,TomitaSA17,DiehlNATPH2008,FinkNatPhys18,FitzpatrickPRX17,GreentreeNatPhys06,Kirton19rev,CarmichaelPRX15}. 
On the experimental side, several platforms have been developed where the physics of non-equilibrium quantum steady states can be studied. One could cite circuit QED \cite{CarusottoNat2020}; arrays of Rydberg atoms \cite{chang2014quantum}, ion traps \cite{Leibfried03} ; and semiconductor microcavities \cite{Kavokin}, only to name the most advanced setups. 
In analogy with equilibrium statistical physics, theoretical understanding often benefits from using ``toy models" able to capture fundamental phenomena. For driven-dissipative systems, the main models under study are Bose-Hubbard type models that are used to describe photonic systems undergoing photon loss  \cite{Hartmannetal2008}; and spin models, that can for instance provide an effective description of Rydberg atoms undergoing spontaneous decay \cite{LeePRL13}. As possible applications of correlations in driven-dissipative lattice-like geometries beyond simulation, we can mention efficient solving of NP-hard optimization problems \cite{Mohseni2022}, related to neural networks \cite{Monroe2014,Ballarini2020,Bravo2022} and metrology \cite{Napolitano2011}, among others.
Even in the case of the simplest models, our theoretical understanding of phase transitions in driven-dissipative systems is hampered both by the absence of a free energy minimisation principle, underpinning the determination of the steady state; and by the difficulty to simulate systems that are sufficiently large to approach the thermodynamic limit.
 In order to address these challenges, a new set of theoretical tools was developed over the last years \cite{Sieberer_2016,DaleyAdvancesinPhysics2014,Weimeretal2021,DeuarPRXQ21,FinazziPRL15, HuybrechtsPRA2020_2, Ramusat2021quantumalgorithm,verstraelenthesis,VerstraelenAS18,mink2022hybrid} for the study of open quantum many-body systems. 
It is to the development of new methods that the present paper wishes to contribute. Here we propose a scheme for dissipative quantum spin lattices, that effectively combines the stochastic sampling of the density matrix evolution by the quantum trajectory method with a scheme of truncation of the correlation-function hierarchy to second order. 
Our method is inspired by approaches already developed in the case of bosonic lattice models \cite{VerstraelenAS18,VerstraelenPRR20,verstraelenthesis}, in which stochastic quantum trajectories are combined with a Gaussian Ansatz on the single-trajectory bosonic state -- the latter state can be viewed as resulting from the Ansatz of vanishing cumulants of the bosonic fields beyond second-order ones. Similarly, we adopt here for the spin operators a truncation of the correlation hierarchy, by assuming that cumulants of order $k$ exceeding a given cutoff $k_c$ vanish -- in practice we take $k_c=2$ in this work, but the same approach can be readily extended to higher-order truncation schemes. The rationale behind the truncation is that higher-order correlations, while developing in closed quantum systems, are far more vulnerable to decoherence in open ones, so that their influence on the dynamics remains limited \cite{Deutsch2020}. 
Our method is sufficiently simple to treat quite large systems comprising hundreds of qubits, and at the same time it allows in principle for a faithful description of long-range quantum correlations, which is a fundamental requirement in order to describe phase transitions that are either driven or significantly altered by quantum effects. 

As an illustration of our approach, we apply our method to the driven-dissipative XYZ model in two spatial dimensions, that can effectively describe the dynamics of an array of optically driven Rydberg atoms~\cite{LeePRL13};  and which has attracted a considerable attention on the theory level as an effective model for driven-dissipative phase transitions \cite{JinPRX16, RotaPRB17, CasteelsPRA18, RotaNJP18, BiellaPRB18, HuybrechtsPRA19, NagyPRL19, HuybrechtsPRB20,Li2020,Kilda2020,McKeever20,Owen2018,ChanPRA2015, LiPRB2021, Jin2022}. It combines anisotropic nearest-neighbour couplings with spin-flip dissipation; and, in spite of its simplicity, it gives rise to a very rich phase diagram, already at the mean-field level \cite{LeePRL13}. Moreover, it has become clear that the mean-field predictions are qualitatively wrong in some parameter regimes \cite{JinPRX16,CasteelsPRA18, HuybrechtsPRA19}, pointing at the fundamental role of correlations, and posing the challenge of their faithful description. 
Here we find that the simplest non-trivial formulation of our method, focusing on two-point quantum correlators, leads to surprisingly good agreement with the exact predictions for small systems; and it shows a clear trend of improvement with increasing system size.  
Using finite-size scaling, we are able to characterize the phase transitions present in the dissipative XYZ system, and confirm that they belong to the 2D classical Ising universality class. This result is to be contrasted with that obtained by simply including classical correlations within a mean-field trajectory scheme (corresponding to $k_c=1$ in the cumulant-truncation approach introduced below), whose results for the transition are inconsistent with the 2D Ising universality class. This observation suggests a surprisingly crucial role of quantum correlations (fully discarded in the $k_c=1$ scheme, and accounted for at all length scales in the $k_c=2$ one) in determining the critical behavior of the open quantum system, in spite of the apparent classical nature of the observed criticality.   
Furthermore, we inspect the nature of quantum correlations and observe that the system develops spin squeezing in the steady state, witnessing short-range entanglement. Moreover we analyze an upper-bound of the quantum Fisher information based on a convex-roof construction \cite{TothP2013}, and show that spin squeezing comes close to saturating this bound in the vicinity of the dissipative phase transition. 

Our paper is structured as follows: in Section \ref{sec:XYZ}, we describe the dissipative XYZ model under study, and review previous numerical approaches used to study this and other dissipative many-body models. In Section \ref{sec:method}, we present the details of the correlation-hierarchy method for dissipative spin systems. Section \ref{sec:results} is devoted to the discussion of the results, and in particular of the finite-size scaling analysis of the dissipative phase transition, as well as of the quantum correlation properties across the phase diagram. Conclusions are offered in Sec. \ref{sec:concl}.

\section{Dissipative quantum spin lattices}
\label{sec:XYZ}

\subsection{The dissipative XYZ model }
We focus our attention on the two dimensional dissipative XYZ Heisenberg model, which has been the subject of intensive theoretical research efforts in the past years \cite{LeePRL13, JinPRX16, RotaPRB17, CasteelsPRA18, RotaNJP18, BiellaPRB18, HuybrechtsPRA19, NagyPRL19, HuybrechtsPRB20,Li2020,Kilda2020,McKeever20,Owen2018,ChanPRA2015, LiPRB2021, Jin2022}, due to its rich phase diagram. Its coherent dynamics is governed by the anisotropic Heisenberg Hamiltonian:
\begin{equation}
    \hat{H} = \sum_{\langle i,j\rangle} \left (  J_x\hat{\sigma}_i^x\hat{\sigma}_{j}^x + J_y\hat{\sigma}_i^y\hat{\sigma}_{j}^y + J_z\hat{\sigma}_i^z\hat{\sigma}_{j}^z \right )~.
    \label{eq:HXYZ}
\end{equation}
Here $J_\alpha$'s are the coupling strengths for the $\alpha = x,\ y, \ z$  spin components of nearest-neighbouring spins; $\hat{\sigma}_i^\alpha$ are the Pauli matrices acting on site $i$; and $\sum_{\langle ij\rangle}$ runs over all pairs of nearest neighbours $\langle ij \rangle$. The XYZ couplings can be engineered in e.g. Rydberg or dipolar atoms, through a combination of dipole interactions with engineered optical pumping \cite{LeePRL13}. Note that energy minimization alone would favor antiferromagnetic order for positive values of the coupling constants, and ferromagnetic order when they are all negative; yet the steady state of the dissipative dynamics challenges the prediction of energy minimization, and it exposes its fundamental non-equilibrium nature. 

The dissipative part of the dynamics stems from spontaneous decay, described as incoherent spin flips along the $z$-axis.  Each spin is coupled to its own Markovian environment, so that the equation governing the dissipative dynamics of the quantum state $\hat \rho$ can be assumed to be of the Gorini-Kossakowski-Sudarshan-Lindblad (GKSL) form \cite{Lindblad1976, Gorini1976, BreuerBookOpen}
\begin{equation}\label{eq:GKSL}
    \partial_t\hat{\rho} = -i\left[\hat{H},\hat{\rho}\right] + \frac{1}{2}\sum_j\left(2\hat{\Gamma}_j\hat{\rho}\hat{\Gamma}_j^\dagger - \hat{\Gamma}_j^\dagger\hat{\Gamma}_j\hat{\rho} - \hat{\rho}\hat{\Gamma}_j^\dagger\hat{\Gamma}_j\right).
\end{equation}
Here, the $\hat{\Gamma}_j = \sqrt{\gamma}\hat{\sigma}_j^-=\sqrt{\gamma} (\sop^x_j-i\sop^y_j)/2$ are the Lindblad operators for the incoherent spin-flip processes along the $z$-axis with dissipation rate $\gamma$. The description of spontaneous decay in quantum optics by the GKSL equation is widely accepted \cite{BreuerBookOpen}, and initiated by Ref. \cite{LeePRL13} for the study of the dissipative XYZ model. We note that in more general open quantum systems, this validity is not always true \cite{Tupkary22}, but it can often be justified through collisional models \cite{Cattaneo21,Barra2015}, especially in cases where external driving is present. 

Along the dynamics governed by the master equation Eq.~\eqref{eq:GKSL}, the expectation value of an operator evolves according to the equation
\begin{align}
\label{eq:masterhierarchy}
\dd{\ev{\Oop}}= & i\ev{\comm{\Ham}{\Oop}} \dd{t}\\ &-\frac{1}{2}\sum_j\left(\ev{\Gamop_j^\dagger\comm{\Gamop_j}{\Oop}}-\ev{\comm{\Gamop^\dagger_j}{\Oop}\Gamop_j}\right)\dd{t},\nonumber
\end{align}
from which the time evolution of the moments of the spin operators can be straightforwardly constructed.

 A pioneering mean-field study of the XYZ model \cite{LeePRL13}, based on the assumption of a fully factorized quantum state $\hat \rho = \otimes_i \hat\rho_i$ at all times, has revealed a very rich phase diagram, containing a paramagnetic phase; a ferromagnetic phase; an antiferromagnetic phase; as well as a spin-density wave and a staggered XY phase. Subsequent works, based on methods going beyond a single-site mean-field approach, offer substantially different predictions for the phase diagram when the XYZ model is cast on a two-dimensional lattice. Cluster mean-field descriptions on the level of the master equation \cite{JinPRX16, BiellaPRB18, Li2020,Jin2021}, which are able to incorporate the influence of some short-range quantum and classical correlations, revealed a drastic impact of such correlations on the phase diagram of the system, a feature not observed in their closed-system counterparts. However, as such they completely miss the long-range fluctuations, which  actually govern the critical behavior. 

On the other hand, the mean-field approximation applied at the level of the quantum trajectory formalism -- namely by studying each trajectory via a Gutzwiller Ansatz -- allows for the inclusion of long-range \emph{classical} correlations \cite{CasteelsPRA18}, possibly combined with some short-range quantum correlations when using a cluster Gutzwiller wave-function \cite{HuybrechtsPRA19}.
Unfortunately in all cluster approaches -- both at the level of the master equation as well as at the level of quantum trajectories -- one is still limited by the relatively small size of the clusters that can be used, and the faithful description of correlations is therefore not guaranteed. Moreover, as we shall see in this work, the critical behavior predicted by Gutzwiller trajectories is very different from from what one can obtain with improved methods, taking into account quantum correlations to all scales. 

It is worth mentioning that the dissipative XYZ model is efficiently solvable in the case of all-to-all connectivity, corresponding to a system in infinite spatial dimensions \cite{HuybrechtsPRB20}. The efficient solution, allowing one to treat exactly systems with hundreds of spins, exploits the permutational symmetry of the system in infinite dimensions; and it has been used to test the validity of the mean field approximation in this limit \cite{HuybrechtsPRB20}. Some extentions to frustrated lattice geometries were studied in \cite{Li2020,LiPRB2021} on the mean-field or cluster mean-field level.
The literature has so far predominantly focused on the transition from the paramagnetic phase to the ferromagnetic phase, which \emph{e.g.} takes place for $J_y \approx \gamma$ when one chooses $J_x = 0.9\gamma$ and $J_z = \gamma$. 

 Using beyond-mean-field methods (see the following section), Refs.~\cite{RotaPRB17,BiellaPRB18,Jin2021} have provided estimates of the critical exponent $\gamma\approx1.5$, which clearly indicates a departure from mean-field prediction. However, the convergence of these results is hard to assess and could therefore either be interpreted as hinting towards a known set of exponents such as the ones of the Ising universality class \cite{RotaPRB17}, or pointing to a new class on its own \cite{Jin2021}.
Extensions to the initial mean-field study in Ref. \cite{LeePRL13}, as well as exact numerical methods in small systems \cite{RotaNJP18}, 
have suggested a reentrant behavior: upon increasing the value of $J_y$ further, ferromagnetic order is lost, and the system reenters into a paramagnetic phase. The study of this second transition and of its critical properties, however, is not a simple task. Interestingly, some of the methods either did not observe the reentrant behavior or suggested that it was not associated with a true phase transition \cite{RotaNJP18,RotaPRB17,NagyPRL19}. Most of the beyond mean-field methods are limited by system size, or do not converge in this region of high $J_y$, where it has been shown that the state is highly mixed \cite{JinPRX16, RotaPRB17, HuybrechtsPRB20}. 

In the following sections we shall see that our method allows us to overcome the above difficulties, and to provide a quantitative analysis of all the transitions of the system on large lattices (comprising up to hundreds of spins), as well as an assessment of the importance of quantum correlations. 

\subsection{Numerical techniques for the study of dissipative spin lattices}

Before ending this section we would like to survey other numerical approaches that have been developed for the study of dissipative quantum spin models in the recent past (see also \cite{Weimeretal2021} for a recent review), and which have been or could be applied to the study of dissipative phase transitions such as those of the XYZ model. 

Several methods have been proposed in order to overcome the limitations of cluster mean-field approaches, based on different Ans\"atze for the density matrix solving the Lindblad master equation. One such method is the corner-space renormalization method \cite{FinazziPRL15, RotaPRB17}, which searches for a solution of the master equation for a given system size within a reduced Hilbert subspace, built from the most likely pure states appearing in the steady state of a smaller system size. Such a method allows one a priori to account for quantum correlations encompassing the whole system; yet, due to the choice of a reduced Hilbert space growing polynomially with system size, it is limited in the amount of entropy that the steady state can exhibit. Recently, variational Ans\"atze based on neural network quantum states \cite{NagyPRL19, HartmannPRL19, VicentiniPRL19, YoshiokaPRB19} have also been proposed, showing promise for the description of large lattices, application to the dissipative XYZ model has not seen the re-entrant paramagnetic behavior \cite{NagyPRL19}. 
Another class of Ans\"atze is offered by tensor network states \cite{Weimeretal2021,DaleyAdvancesinPhysics2014}: within this approach, infinite projected entangled pair states (iPEPS) are particularly relevant, as they immediately grant access to the thermodynamic limit in the regions where they converge \cite{Kilda2020, McKeever20}. Application to the dissipative XYZ model has been considered in \cite{Kshetrimayum17}. It has been found however that this approach is subject to instabilities \cite{Kilda2020}. Attempts to overcome this issue have been suggested \cite{McKeever20,Kilda2020}, but have to our knowledge not been tested to the general dissipative XYZ-model. Generally speaking, tensor network states have shown to be most effective for the study of 1D systems \cite{DaleyAdvancesinPhysics2014}. 

Finally, an alternative category of methods for dissipative spin lattices relies on the truncated Wigner approximation (TWA) \cite{SinghW2022,Huberetal2022,mink2022hybrid}, which represents the expectation values of operators along the quantum dynamics as averages over evolutions of classical spins, starting from states drawn out of the discrete Wigner distribution for the quantum spins. This approach can tackle very large lattice sizes, and effectively account for classical correlations, but it has the disadvantage that the quantum nature of the system is mimicked by classical fluctuations
and it is therefore unclear whether quantum effects are properly described in the steady state. For example, the TWA has been shown to suffer from unphysical predictions for the single-photon driven Bose-Hubbard model \cite{Regemortel17}  and also to miss the phase transition in the steady state under two-photon driving \cite{verstraelenthesis}, motivating the construction of techniques that are complementary to the TWA.

\section{Correlation hierarchies for dissipative systems \label{sec:method}}

\subsection{Open quantum systems: quantum trajectories, classical vs. quantum fluctuations \label{sec:oqs}}

\subsubsection{Stochastic unraveling of the open-system evolution}

The most fundamental level of description for the dynamics of quantum systems coupled to a Markovian bath is offered by the GKSL equation Eq.~\eqref{eq:GKSL} \cite{BreuerBookOpen,Gardiner_BOOK_Quantum,Carmichael_BOOK_2,Wiseman_BOOK_Quantum,simplemodel} for the evolution of the density matrix of the system. Nonetheless the theoretical framework which we shall use here is that of quantum trajectories \cite{DaleyAdvancesinPhysics2014}, which stochastically sample the density matrix. 
In order to understand this framework, one should realize that the master equation Eq.~\eqref{eq:GKSL} is largely independent of the environment specifics. We may thus replace the true environment with a macroscopic measuring device without an observer.
In practical terms, this means that the incoherent effect of the environment is the same as the one that would arise from a continuous weak, non-selective measurement. It is precisely because the measurement is not read out, that classical uncertainty over the system increases and Eq. \eqref{eq:GKSL} does not conserve purity. But one may also consider what happens if one does observe the state of the measuring device. The time-evolution is then conditioned on the measurement result, and, assuming no measurement imperfections, there is no loss of information, so that states remain pure. Such a conditional evolution consists of \emph{quantum trajectories}, that were introduced in the seminal works of Refs.~ \cite{carmichael_1993, DalibardPRL92, Dum92, Barchielli_1991}. 
The expectation values associated with the density matrix evolved with the master equation Eq.~\eqref{eq:GKSL} are then recovered by averaging over the pure states of the trajectories (that is, tracing out the information on the measurement outcome), in the limit of a sufficiently large sample. 
This latter approach is known in the literature as wavefunction Monte Carlo, or the stochastic sampling method. The advantage of wavefunction Monte Carlo, compared to the direct solution of the master equation, is that the pure states of the trajectories can be described by a wavefunction $\ket{\psi}\in\mathcal{H}$ with $D = {\rm dim}({\cal H})$ components only, compared to a density matrix $\Rho\in\mathcal{H}\otimes\mathcal{H}^*$ with $D^2$ components, so that significantly less computer memory is required for the exact numerical simulation of trajectories. 

Within this picture of quantum trajectories, there is still some freedom regarding the (possibly hypothetical) measurement protocol that is performed, defining a so-called \emph{unraveling} scheme. As long as no additional approximations are made, the result of averaging over trajectories should be consistent with the master equation independently of the measurement protocol, even though the individual trajectories may have a vastly different behavior depending on the measurement protocol itself. 

One of the most adopted unraveling schemes is based on quantum jumps (which models \emph{e.g.} photon counting in experiments on optical cavities), leading to trajectories of the form 
\begin{align}\label{eq:pctrajeq}
    \dd{\widetilde{\ket{\psi}}}=-i \Ham_{\text{traj}} \widetilde{\ket{\psi}} \dd{t},
\end{align}
where 
\begin{equation}
\Ham_{\text{traj}}=\Ham - \frac{i}{2}\sum_j \Gamop^\dagger_j \Gamop_j,
\end{equation}
is an effective non-Hermitian Hamiltonian. Equation \eqref{eq:pctrajeq} does not conserve the norm of the wavefunction, hence the tilde-notation. This evolution is complemented by the fact that, given a uniform stochastic random number $0<\zeta<1$ , whenever $\norm{\widetilde{\ket{\psi}}}^2<\zeta$, a discrete jump of the form
\begin{equation}\label{eq:pctrajeq2}
\widetilde{\ket{\psi}}\jump\frac{\Gamop_l \widetilde{\ket{\psi}}}{\norm{\Gamop_j \widetilde{\ket{\psi}}}},    
\end{equation}
will take place, where the choice of $\Gamop_l$ has probability $\norm{\Gamop_l \widetilde{\ket{\psi}}}^2/\sum_j\norm{\Gamop_j \widetilde{\ket{\psi}}}^2$.
In the case of quantum spin models with spontaneous decay, such a scheme corresponds to having a photon counting apparatus coupled to each spin separately, measuring the number of photons spontaneously emitted by the spin.

A different type of unravelling is obtained by a different measurement scheme, which in quantum optics corresponds to heterodyning -- namely, mixing the emitted photons with a classical field at a different frequency. For heterodyne detection, the single-trajectory evolution is governed by a stochastic Schr\"odinger equation in the form
 \begin{align}\label{eq:hettraj}
 \dd{\ket{\psi}}&=-i\Ham\dd{t}\ket{\psi} \\
 &+\sum_j\left(\ev{\Gamop_j^\dagger}\Gamop_j-\frac{1}{2}\ev{\Gamop_j^\dagger}\ev{\Gamop_j}-\frac{1}{2}\Gamop_j^\dagger\Gamop_j \right)\dd{t} \ket{\psi} \nonumber \\ &+\sum_j\left(\Gamop_j-\ev{\Gamop_j}\right)\dZ{j}^*\ket{\psi},  \nonumber
 \end{align}
where $\dZ{j}=(\dd{W}_j^X+i\dd{W}_j^Y)/\sqrt{2}$ is complex Wiener noise in the Ito sense, satisfying $\dZ{i}^*\dZ{j}=\delta_{ij}\dd{t}, \dZ{i}\dZ{j}=0$. Such an evolution is often referred to as quantum state diffusion \cite{Gisin1992}.
For the evolution of expectation values, Eq. \eqref{eq:hettraj} yields
 \begin{align}\label{eq:hettraj-exp}
\dd{\ev{\Oop}} & =i \ev{\comm{\Ham}{\Oop}}\dd{t} \\
&-\frac{1}{2}\sum_j\left(\ev{\Gamop_j^\dagger\comm{\Gamop_j}{\Oop}}-\ev{\comm{\Gamop_j^{\dagger}}{\Oop}\Gamop_j}\right)\dd{t} \nonumber \\ 
&+\sum_j \left(\ev{\Gamop_j^\dagger (\hat O -\langle \hat O \rangle)}\dZ{j}
+ \ev{(\hat O -\langle \hat O \rangle) \Gamop_j}\dZ{j}^*\right). \nonumber
\end{align}
Comparing this expression with Eq.~\eqref{eq:masterhierarchy}, we see that the time evolution for expectation values under the master equation is identical to the deterministic part of the evolution under the heterodyne quantum trajectories. The stochastic nature of the detector clicks under quantum trajectory evolution reflects itself in the addition of noise to the time evolution of expectation values. 

\subsubsection{Classical vs. quantum fluctuations and correlations}
\label{sec:classquantfluct}

Given the correspondence between the master equation and its stochastic unraveling, one has the choice to choose an Ansatz at either level.
We have already pointed out above the advantage of quantum trajectories in reducing the computational cost for exact calculations. 
The advantage of quantum trajectories is even more dramatic when an Ansatz is made on the state of the system. 
Indeed, unlike the case of an Ansatz for the the density matrix, an Ansatz formulated at the trajectory level is always complemented by classical (\emph{i.e.} trajectory-to-trajectory) fluctuations \cite{DaleyAdvancesinPhysics2014}. Denoting with $|\psi_n(t)\rangle$ the solution of the stochastic Schr\"odinger's equation along the $n$-th trajectory, we shall define as classical fluctuations of an observable $\hat{O}$ its trajectory-to-trajectory fluctuations
\begin{equation}
{\cal F}_{\rm c}(\hat O) = [O^2_n(t)]_{\rm traj} -  [O_n(t)]^2_{\rm traj},
\end{equation}
where $O_n(t) =  \langle \psi_n(t) | \hat O  | \psi_n(t) \rangle$ is the single-trajectory expectation value of the observable; 
$[...]_{\rm traj} = N_{\rm traj}^{-1} \sum_n (...)$ denotes the average over $N_{\rm traj}$ trajectories; and the time $t$ is chosen to be in the regime of convergence to the steady state. The above fluctuations are uniquely due to the effect of the environment, and they correspond to the incoherent part of the fluctuations associated with the pure-state decomposition of the density matrix 
\begin{equation}
\rho(t) = \frac{1}{N_{\rm traj}} \sum_n |\psi_n(t)\rangle \langle \psi_n(t) |,
\end{equation}
 valid asymptotically in the limit $N_{\rm traj} \to \infty$. On the other hand, quantum fluctuations in this context can be defined as the fluctuations proper to the single-trajectory wavefunctions $|\psi_n(t)\rangle$, averaged over all trajectories:
\begin{equation}
{\cal F}_{\rm q}(\hat O) = [\ev{ O^2}_n(t) - \ev{O(t)}_n^2]_{\rm traj},
\label{eq:quantfluct}
\end{equation}
where $\ev{O^2}_n(t) =  \langle \psi_n(t) | \hat O^2  | \psi_n(t) \rangle$. These fluctuations are clearly most sensitive to the choice of the Ansatz for the trajectory wavefunctions. It is easy to verify that the total fluctuations associated with the density matrix can be decomposed into a classical and a quantum part 
\begin{equation}
{\cal F}(\hat O) = \langle \hat O^2 \rangle_{\hat \rho} -  \langle \hat O\rangle_{\hat \rho}^2=   {\cal F}_{\rm c}(\hat O) + {\cal F}_{\rm q}(\hat O)~.
\end{equation}
where $\langle ... \rangle_{\hat \rho} =  {\rm Tr}(\hat \rho ...) $~. 
Note that the decomposition of the density matrix into pure states (and thus relative contributions of $\mathcal{F}_c$ and $\mathcal{F}_q$ ) is not unique, as it depends in principle on the stochastic unraveling that one chooses. 

A similar decomposition can be carried out in the case of correlations, namely, given two local operators $\hat A$ and $\hat B$ (for instance $\hat A = \hat \sigma_i^\mu$ and $\hat B = \hat \sigma_j^\nu$ for two local spin components)  
\begin{equation}
C(\hat A,\hat B) = \langle \hat A \hat B \rangle_{\hat \rho} - \langle \hat A \rangle_{\hat \rho}  \langle \hat B \rangle_{\hat \rho} = C_c(\hat A,\hat B) + 
 C_q(\hat A,\hat B),
\end{equation}
where 
\begin{equation}
C_c(\hat A,\hat B)  = [A_n(t) B_n(t)]_{\rm traj} -  [A_n(t)]_{\rm traj} [B_n(t)]_{\rm traj},
\label{eq:classcorr}
\end{equation}
are the classical correlations associated to trajectory-to-trajectory fluctuations; while 
\begin{equation}
C_q(\hat A,\hat B)  = [\langle AB\rangle_n(t) - A_n(t) B_n(t)]_{\rm traj} ,
\label{eq:qcorr}
\end{equation}
are referred to as the quantum correlations, originating from the entangled nature of the wavefunction of each trajectory. 

The above discussion suggests the ability of the stochastic unraveling to capture the essential traits of the classical, incoherent fluctuations and correlations of the density matrix solution to the master equation, regardless of the choice of the Ansatz wavefunction describing each trajectory (but the results of this work shall cast some doubts on this point of view, see Sec.~\ref{sec:phasetrans_expon}).
This aspect has been revealed by several quantum trajectory calculations based on the Gutzwiller Ansatz \cite{CasteelsPRA17dimer,CasteelsPRA18,PichlerPRA10,PichlerPRA13,DiehlPRL10}, the cluster Gutzwiller Ansatz \cite{HuybrechtsPRA19, HuybrechtsPRA2020_2}, the Gaussian Ansatz \cite{VerstraelenAS18,VerstraelenPRR20} and the matrix-product state Ansatz \cite{DaleyPRA9,BarmettlerPRA11}, generally providing superior results with respect to calculations based on similar Ans\"atze formulated at the level of the density matrix. 
 As an example, a Gaussian Ansatz for the trajectory wavefunctions of a bosonic system is able to describe phenomena such as optical bistability in dissipative non-linear cavities that cannot be described by applying the same Gaussian Ansatz at the level of the master equation \cite{VerstraelenAS18,verstraelenthesis}. Indeed the Gaussian Ansatz at the level of the master equation approximates the density matrix to be a single Gaussian state, fully described in terms of the average field quadratures and of their covariance matrix; while the Gaussian Ansatz for trajectory wavefunctions approximates the density matrix as a pure-state decomposition on Gaussian states, which is no longer a Gaussian state -- namely its higher-order correlations for the field quadratures are no longer reducible to one- and two-point ones. 
 
In this work we apply a similar insight to the case of quantum spin systems. The analog of a Gaussian state for spin system is a state whose higher-order spin-spin correlations are reducible to single-spin and two-spin expectation values, assuming the vanishing of all multivariate cumulants for quantum-spin operators beyond the second-order ones. We shall apply such an Ansatz to single trajectories for the study of the evolution of dissipative quantum spin lattices in the next section. Before doing so, nonetheless, we would like to elaborate further on the link between quantum fluctuations and correlations extracted from specific unraveling schemes, and quantum fluctuations and correlations associated with the full state $\hat\rho$ which is stochastically reconstructed by the unraveling scheme. This link is discussed in the next paragraph.
     

\subsubsection{Quantum correlations from trajectories vs. the quantum Fisher information of the full state}
\label{sec:bounds}

The previous paragraph showed that the quantum trajectory approach can shed light on the role of classical vs. quantum fluctuations and correlations in the state of the system, via the combined analysis of the statistics of trajectory-to-trajectory fluctuations vs. that of the fluctuations within single-trajectory wavefunctions. This analysis is clearly dependent on the specific unraveling scheme; and it appears at first sight to be of purely theoretical interest, given that the single-trajectory fluctuations are inaccessible experimentally, unless one is able to repeat the exact same trajectory (i.e. the same sequence of measurement records) multiple times. Nonetheless, the theoretical estimate of quantum fluctuations and correlations at the level of a specific unraveling scheme turns out to be much more relevant than what may at first appear, thanks to the link between this estimate and fundamental quantum coherence properties of the state of the system $\hat \rho$.

A central quantity for the determination of quantum-fluctuation and quantum-correlation properties of a generic state $\hat \rho$ is the quantum Fisher information (QFI) related to an operator $\hat A$ \cite{Braunstein94,PezzeRMP2018}, which is defined as
\begin{equation}
    \text{QFI}(\hat A) = 2\sum_{lm} \frac{(\lambda_l - \lambda_m)^2}{\lambda_l + \lambda_m}\vert\langle l\vert \hat A \vert m\rangle\vert^2,
    \label{eq:QFI}
\end{equation}
where $\vert l\rangle$ and $\vert m\rangle$ are eigenstates of $\hat{\rho}$ with respective eigenvalues $\lambda_l$ and $\lambda_m$. The QFI represents the most important quantity in quantum interferometry, determining the ultimate sensitivity of the state $\hat \rho$ to a unitary transformation generated by $\hat U(\theta) = e^{-i \theta \hat A}$. If $\hat A$ and $\hat \rho$ commute, such sensitivity is zero: hence the QFI ultimately probes the non-commutativity between  $\hat A$ and  $\hat \rho$, or, more explicitly, the amplitude of the quantum fluctuations of $\hat A$ in the state $\hat \rho$.

The QFI can also be used to probe entanglement in the state $\hat \rho$. If $\hat A$ is a sum of local qubit observables, $\hat A = \sum_i \hat A_i$, the state sensitivity to the unitary transformation exceeds the standard quantum limit (SQL) $\Delta\theta = 1/\sqrt{N}$ of independent qubits only if the state is entangled \cite{PezzePRL2009, PezzeRMP2018}, as detected by the fact that the QFI density exceeds unity: $\text{QFI}(\hat A)/N > 1$. In particular, the condition $\text{QFI}(\hat A)/N > p$ (with $p \in \mathbb{N}$) reveals that the system contains $(p+1)$-partite entanglement \cite{TothPRA2012, HyllusPRA2012}. 

In quantum trajectory calculations we do not have direct access to the QFI, which generically requires the full knowledge of the density matrix. The stochastic unraveling of the density matrix evolution  does offer a pure-state decomposition of the state $\hat \rho$, but not its eigenstate decomposition (which is instead required by the definition of the QFI, Eq.~\eqref{eq:QFI}). Nonetheless the knowledge of a pure-state decomposition of the density matrix still gives access to an \emph{upper} bound to the QFI \cite{TothP2013,tothuncertainty2022}, in the form of the quantum fluctuations defined in Sec.~\eqref{sec:classquantfluct}, namely
\begin{equation}
{\rm QFI}(\hat A)  \leq 4\sum_n p_n \text{Var}_{\psi_n}(\hat A) = 4 F_q(\hat A) ~.
\label{eq:upperbound}
\end{equation}
Here $p_n$ is the probability assigned to the pure state $|\psi_n\rangle$ by the quantum trajectory approach (\emph{e.g.} in the steady state), and  $ \text{Var}_{\psi_n}(\hat A) = \langle \psi_n | \hat{A}^2 |  \psi_n \rangle  - \langle \psi_n | \hat A |  \psi_n \rangle^2$ is the variance of  $\hat A$ on the state of a single trajectory; hence this expression is equivalent to the trajectory-sampling expression Eq.~\eqref{eq:quantfluct} for the quantum fluctuations in the limit $N_{\rm traj} \to \infty$. 
The above inequality Eq.~\ref{eq:upperbound} becomes an equality only when extremizing $F_q$ over all pure-state decompositions  \cite{TothP2013,tothuncertainty2022}.
Given that the value of the upper bound in Eq.~\ref{eq:upperbound} is unravelling-dependent, one could in principle find an optimal upper bound by exploring different unravellings, although we limit ourselves to heterodyne unravelling for this work.

The expression of the quantum fluctuations of $\hat A$ for the trajectory approach can be made more explicit by casting it in terms of quantum correlations $C_q$ defined in Eq.~\eqref{eq:qcorr}:
\begin{equation}
    F_q(\hat A) = \sum_{i,j} C_q(\hat A_i, \hat A_j).
\end{equation}
Such a decomposition can be put in parallel with that of the QFI for the macroscopic observable $\hat{A}$
\begin{equation}
{\rm QFI}(A) = \sum_{ij} {\cal Q}_{ij},     
\end{equation}
where ${\cal Q}_{ij} = {\cal Q}(\hat A_i, \hat A_j)$ is given by 
\begin{equation}
{\cal Q}_{ij} = 2\sum_{lm} \frac{(\lambda_l - \lambda_m)^2}{\lambda_l + \lambda_m} \langle m | \hat A_i | l \rangle \langle l | \hat A_j |m\rangle,
\end{equation}
and it is the so-called Quantum Fisher information matrix - QFIM (for the special case of commuting observables $[\hat A_i, \hat A_j]=0$), playing a central role in multiple phase estimation \cite{Paris2009, Safranek2017}.

As we shall see for the case of the dissipative XYZ model studied below, the unraveling dependent quantum correlation function $C_q(\hat A_i, \hat A_j)$ has an exponentially decaying behavior as a function of the distance in the steady states that we shall examine below $|C_q(\hat A_i, \hat A_j)| \sim \exp(-r_{ij}/\tilde\ell_Q)$, where $\tilde \ell_Q$ is an (unraveling dependent) \emph{quantum coherence length}, introduced for equilibrium mixed states in Ref.~\cite{MalpettiR2016}, and defining the characteristic spatial range of quantum correlations. Notably, this behavior is also generally expected for thermal states \cite{Hauke2016,MalpettiR2016,Frerotetal2022}, as recently proven in Ref.~\cite{Saito2022}. It is reasonable to assume as well that the QFIM has a similar exponential decay, $|{\cal Q}_{ij}| \sim \exp(-r_{ij}/\ell_Q)$, where now the quantum coherence length $\ell_Q$ is an absolute property of the state $\hat \rho$ and not of its stochastic sampling.  The fact that the integral of $C_q$ (given by $F_q$) is an upper bound to the integral of the QFIM (the QFI itself) leads one to conclude that $\tilde \ell_Q \gtrsim \ell_Q$, namely the spatial range of the unraveling-dependent quantum correlations $C_q$ provides in practice an upper bound to that of the spatial range of the QFIM.

Hence the above inequalities show that the quantum-trajectory analysis of quantum fluctuations provides upper bounds to the QFI and quantum coherence length. 
One may expect the inequalities to be rather tight, since the unraveling corresponding to quantum state diffusion contains local jump operators only, and it is therefore likely to feature close-to-minimal entanglement in the trajectory states \cite{regemortel22}.  In Sec.~\ref{sec:squeezing} we will also provide a lower bound to the QFI, offered by the spin-squeezing parameter, allowing us to provide a definite quantitative estimate of the QFI whenever the lower and upper bounds are close to each other.

\subsection{Cumulant expansion and its truncation}

In analogy with a classical probability distribution, the quantum state of a lattice system can either be specified by its expansion coefficients with respect to a Hilbert space basis; or by the expectation values of a suitable set of single-site operators $\hat X_i$ (e.g. creation and annihilation operators for bosonic and fermionic systems, Pauli matrices for spin systems). The moments of the local operators $\hat X_i$ are expectation values of the form $\langle \hat X_i^m \hat X_j^n \cdots\rangle$, where $k=m+n+\cdots$ is the order of the moment. Such moments can be conveniently expressed in terms of the multivariate \emph{cumulants} 
$\langle \hat X_i^m \hat X_j^n \cdots\rangle_c$, which are recursively defined as \cite{FRICKE1996479,Colussi2020}
\begin{align}
&\langle \hat X_i \rangle =  \langle \hat X_i \rangle_c \nonumber \\
&\langle \hat X_i \hat X_j \rangle  =  \langle \hat X_i \hat X_j \rangle_c + \langle \hat X_i \rangle_c \langle \hat X_j \rangle_c \nonumber \\
&\langle \hat X_i \hat X_j \hat X_k \rangle  =  \langle \hat X_i \hat X_j \hat X_k \rangle_c \nonumber \\ 
&+ \langle \hat X_i \hat X_j \rangle_c \langle \hat X_k \rangle_c  + \langle \hat X_i \hat X_k \rangle_c \langle \hat X_j \rangle_c + \langle \hat X_j \hat X_k \rangle_c \langle \hat X_i \rangle_c \nonumber \\
&+ \langle \hat X_i \rangle_c \langle \hat X_j \rangle_c \langle \hat X_k \rangle_c  \\
&\ldots \nonumber
\end{align}

A crucial insight justifying the use of a cumulant expansion is that, in typical situations of interest in physics -- e.g., in the equilibrium state of many-particle systems -- the value of the cumulants is expected to decrease with their order $k$. Therefore a meaningful approximation scheme might consist in truncating the cumulant hierarchy to a given order $k_c$: in so doing, moments of order $k>k_c$ can be expressed in terms of the moments of order $1 \leq k \leq k_c$, and therefore the whole state is assumed to be described in terms of a finite set of moments, whose number grows polynomially with system size as $N^{k_c}$. 
Approaches based on a truncation of the cumulant expansion for the moments of the fluctuations have been widely used in complex systems across physics: in condensed matter and chemical systems \cite{Colussi2018,FRICKE1996479,Caruso2020,SnchezBarquilla2020} 
but also quantum chromodynamics \cite{OzonderPRD17}; cosmology \cite{Erschfeld2020}; and even medical imaging \cite{MOHANTY201880}.
In the context of driven-dissipative systems, cumulant approaches have been most commonly applied to bosonic systems and in particular on the level of the master equation,
\cite{Rudiger1990,Leymann2014,CasteelsNJP16,Liew2011,Regemortel17,Tonielli2019,Kirton17,Plankensteiner21,Robicheaux21,Marino2019}. Some recent work has shown advantage of descriptions on the level of quantum trajectories for such systems instead, in particular at the Gaussian level \cite{VerstraelenAS18,VerstraelenPRR20,Graefe22}. Our goal here is to extend these methods to dissipative quantum spin lattices, exhibiting their potential to quantitatively describe dissipative phase transitions and the role of quantum fluctuations. 


It must be noted that, while the truncation of the cumulant hierarchy leads often to insightful results, it is typically not variational: namely, it does not necessarily correspond to an existing Ansatz for the quantum state. According to a well-known theorem of statistics proven by Marcinkiewicz \cite{Marcinkiewicz1939}, all classical probability density functions either have only nonvanishing $k\leq2$ cumulants, or cumulants up to infinitely high order will be nonvanishing \cite{kenney_keeping_1961}. This has since been generalized to bosonic quantum states \cite{Rajagopal1974}. 
As a consequence of the Marcinkiewicz theorem, physical bosonic states have either only first-order cumulants (coherent states, $k_c=1$), or only first- and second-order cumulants (Gaussian states, $k_c=2$), or otherwise have non-zero cumulants to all orders. 
For the case of a spin systems, we are not aware of an analog to Marcinkiewicz theorem, although the existence of spin-to-boson mappings suggests that the limitations on the existence of physical truncations of the cumulant expansion valid for bosons may have similar repercussions on spins as well. We argue nonetheless that, even if the truncation to order $k_c$ of the cumulant hierarchy did not correspond to any physical state for the quantum spins, its application to calculations amounts to an embedding of the physical problem of interest within a larger family of problems, a procedure which is rather common in physics. As an example, in the context of spin-to-boson mappings, quantum spins are mapped onto bosons with a constrained Hilbert space \cite{Auerbach1994}, but the necessity to release the constraints for the sake of feasible calculations embeds the quantum-spin problem within a larger family of problems. The latter act is meaningful as long as the physical content of the theoretical predictions is not substantially altered by the embedding. In the following, we shall make this (falsifiable) assumption of the truncation of the cumulant expansion to order $k=k_c=2$ for spin systems. We would like to remark that throughout our study we have not encountered a single unphysical result justifying the need to revise this assumption, and have validated consistency with exact results where these could be obtained.

\subsection{Truncated cumulant equations for dissipative spin systems}\label{sec:cumapproachspinsyst}

In this work we shall focus our attention on spin-1/2 spins, in relationship with current studies on ensembles of qubits coupled to each other and with an environment. 
For a spin-1/2 system, the local spin variables form a closed algebra, such that any product of spin operators acting on the same site can be written as a single spin variable. Therefore the only local observables $X_i$ of interest are the spin components $\hat \sigma_i^\alpha$, $\alpha = x, y, z$, taken to first power, whereas the nonlocal moments of interest are of the kind $\langle \hat \sigma_i^{\alpha_i} \hat \sigma_j^{\alpha_j} \hat \sigma_k^{\alpha_k} ...\rangle $, with $i\neq j\neq k$.
As a consequence, a truncation of the cumulant hierarchy to order $k_c$ implies that all $k$-point correlation functions with $k>k_c$ can be expressed in terms of $k$-point ones with $k\leq k_c$. In the following we shall adopt the truncation scheme of the cumulant expansions at the level of single trajectories, with $k_c=2$. We note that some recent works have applied a similar truncation scheme to spin systems \cite{Plankensteiner21,Robicheaux21}, but as an Ansatz on the density matrix solving the master equation instead. 
In view of the discussion provided above on classical vs. quantum correlations, we can state that the above-cited works truncate all correlations equally, whereas the approach we discuss here preserves classical fluctuations up to all orders, and only truncates the higher-order cumulants of quantum correlations. This is a much more flexible Ansatz since, in analogy with the bosonic case, cumulants of order higher than $k=2$ are preserved at the level of the trajectory-to-trajectory fluctuations.
  We note that a truncation on quantum correlations is physically justified in a dissipative system, as high-order cumulants of quantum fluctuations may be expected to be strongly suppressed by decoherence \cite{Deutsch2020}, given that each site is assumed to interact with its own independent environment.

In the following we shall discuss the single-trajectory equations of motion for the moments of the spin fluctuations within the two truncation schemes with $k_c=1$ (corresponding to the Gutzwiller mean-field Ansatz for trajectories) and with $k_c=2$, which is the truncation level we adopted for the rest of our work. 

\subsubsection{$k=1$ truncation\label{subsubsec:k1}}

When truncating the cumulant expansion to first-order cumulants, nonlocal quantum correlations are completely discarded, and the pure state along each trajectory corresponds to a factorized Gutzwiller Ansatz  \cite{CasteelsPRA18}.
We then obtain from Eqs. \eqref{eq:HXYZ}, \eqref{eq:GKSL} and \eqref{eq:hettraj}
 \begin{widetext}
 \begin{align}
    \dd{\es{m}{x}}=&\left(-\frac{\gamma}{2}\es{m}{x}+2J_y \sum_{m'} \es{m'}{y}\es{m}{z}-2J_z \sum_{m'} \es{m'}{z}\es{m}{y}\right)\dd{t}\nonumber\\&+\sqrt{\frac{\gamma}{2}}\left(1+\es{m}{z}-\es{m}{x}^2\right) \dd{W}^x_m+\sqrt{\frac{\gamma}{2}}\es{m}{x}\es{m}{y} \dd{W}^y_m \label{eq:k1x}\\
    \dd{\es{m}{y}}=&\left(-\frac{\gamma}{2}\es{m}{y}+2J_z \sum_{m'} \es{m'}{z}\es{m}{x}-2J_x \sum_{m'} \es{m'}{x}\es{m}{z}\right)\dd{t}\nonumber\\&-\sqrt{\frac{\gamma}{2}}\left(1+\es{m}{z}-\es{m}{y}^2\right) \dd{W}^y_m-\sqrt{\frac{\gamma}{2}}\es{m}{x}\es{m}{y} \dd{W}^x_m  \label{eq:k1y}\\
    \dd{\es{m}{z}}=&\left(-\gamma (\es{m}{z}+1)+2J_x \sum_{m'} \es{m'}{x}\es{m}{y}-2J_y \sum_{m'} \es{m'}{y}\es{m}{x}\right)\dd{t}\nonumber\\
    &-\sqrt{\frac{\gamma}{2}}\es{m}{x}(1+\es{m}{z})\dd{W}^x_m+\sqrt{\frac{\gamma}{2}}\es{m}{y}(1+\es{m}{z})\dd{W}^y_m  \label{eq:k1z}~.
\end{align}
 \end{widetext}
 
Please notice that these equations are expressed differently than the ones from Ref.~\cite{CasteelsPRA18}, but are equivalent (modulo the choice of stochastic unraveling). We have also observed that they are numerically more efficient to integrate. When the noise terms are omitted in Eqs. \eqref{eq:k1x}-\eqref{eq:k1z}, recovering the $k=1$ truncation scheme at the level of the density matrix, these equations  reduce to the mean-field equations  described in Ref.~\cite{LeePRL13}. 

 In Appendix~\ref{ap:k1} we extend the $k=1$ calculations of Ref.~\cite{CasteelsPRA18}, and we provide a comprehensive finite-size-scaling analysis of the dissipative paramagnetic-ferromagnetic phase transition exhibited by the $k=1$. 
This analysis shows that the $k=1$ data are incompatible with the universality class of the 2D classical Ising, but have rather good agreement with the mean field universality class; however, this picture is strongly altered by the inclusion of quantum correlations within the $k=2$ truncation scheme, as we shall further discuss in Sec.~\ref{sec:phasetrans_expon}, which is in stark contrast with the treatment of classical correlations only.

\subsubsection{$k=2$ truncation}

We now turn our attention to the more general $k>1$ case, which allows one to account for quantum correlations in the dissipative dynamics. The derivation of the equations for the evolution follows the scheme outlined above, and in the $k=2$ case it is detailed in Appendix \ref{sec:hetGauss} for heterodyne unraveling.

As already mentioned above, in the bosonic case only $k=1,2$ truncations correspond to variational states (coherent states and Gaussian states respectively). 

In the case of spin states, the situation is more intricate. The vanishing of cumulants of order larger than $k=1$ is realized the by Gutzwiller states (see \ref{subsubsec:k1}). The absence of well understood quantum spin states that display closure at the $k = 2$  level, however, does not mean that this truncation is without its merits in the description of physical systems.

It has actually been shown that almost all pure quantum states for $N=3$ spins are in fact fully determined by the knowledge of the two-spin density matrices \cite{Linden2002}. This implies in turn that third-order cumulants must vanish exactly for these states. This result has been generalized to systems with $N>3$ \cite{Linden2002bis,Zhou2008}, where it has been proven that most quantum states are fully determined by the knowledge of the reduced states involving only a fraction of its degrees of freedom -- albeit generically a macroscopic fraction thereof. This suggests that truncated cumulant hierarchies have physical relevance, even though the truncation to order $k_c=2$ might not be realized exactly by a generic physical state for $N\gg2$ spins.

The choice of $k=2$ is the one guaranteeing the ability to capture quantum correlation at the smallest computational cost -- namely, that of tracking the evolution of 2-point correlation functions, requiring a computational time of order ${\cal O}(N^2)$ for systems with short-range interactions. As noted above, these two-point quantum correlation functions are the ones most likely to survive decoherence and they are expected to dominate the critical behaviour at true quantum phase transitions.

 A technical remark is in order at this point. In the stochastic trajectory approach, the fluctuations of order $\sqrt{\dd{t}}$ result in equations that turn out to be numerically unstable. Stability is recovered however when the noise terms in the equations of motion for the second-order cumulants are dropped. The justification for omitting the noise terms in the dynamics of the second-order cumulants is detailed in Appendix \ref{ap:measeff}. There we show that the dynamics becomes more stable when the amplitude of the noise is reduced. Physically, this corresponds to a situation with detectors that have a limited efficiency, namely that they only detect a portion 
of the photon signal emitted by the spins upon decaying. With these finite efficiencies, we see that the presence of the noise on the second moments does not affect the results, and we can extrapolate the results to the (numerically unstable) trajectory limit  with $\gamma_m=\gamma$. This observation allows us therefore to omit these noise terms on the second-order cumulants. Doing so,
 we observe very good agreement between the cumulant hierarchy and numerically exact results for small systems -- as we shall detail below -- further demonstrating the validity of this approach.

\section{Results \label{sec:results}}
In what follows we present the results for the dissipative XYZ model from our method obtained with the $k = 2$ cumulant truncation. 
In Secs.~\ref{sec:structfact}, \ref{sec:pm_quant}, \ref{sec:corrfunc}, \ref{sec:phasetrans_expon} and \ref{sec:squeezing}, we shall focus our attention on the case $J_x = 0.9\gamma$, and $J_z = \gamma$; a full account of the phase diagram of the system will be provided in Sec.~\ref{sec:phd}. Note that to obtain the expectation value of an operator $\hat{O}$ in the quantum trajectory formalism, we will time evolve each quantum trajectory over a long enough period of time for it to reach the steady state regime. Subsequently, we continue to time evolve and use this collected data to perform time averaging when calculating expectation values. Additionally, we average over multiple trajectory realizations.

\begin{figure}
    \centering
    \includegraphics[width=0.5\textwidth]{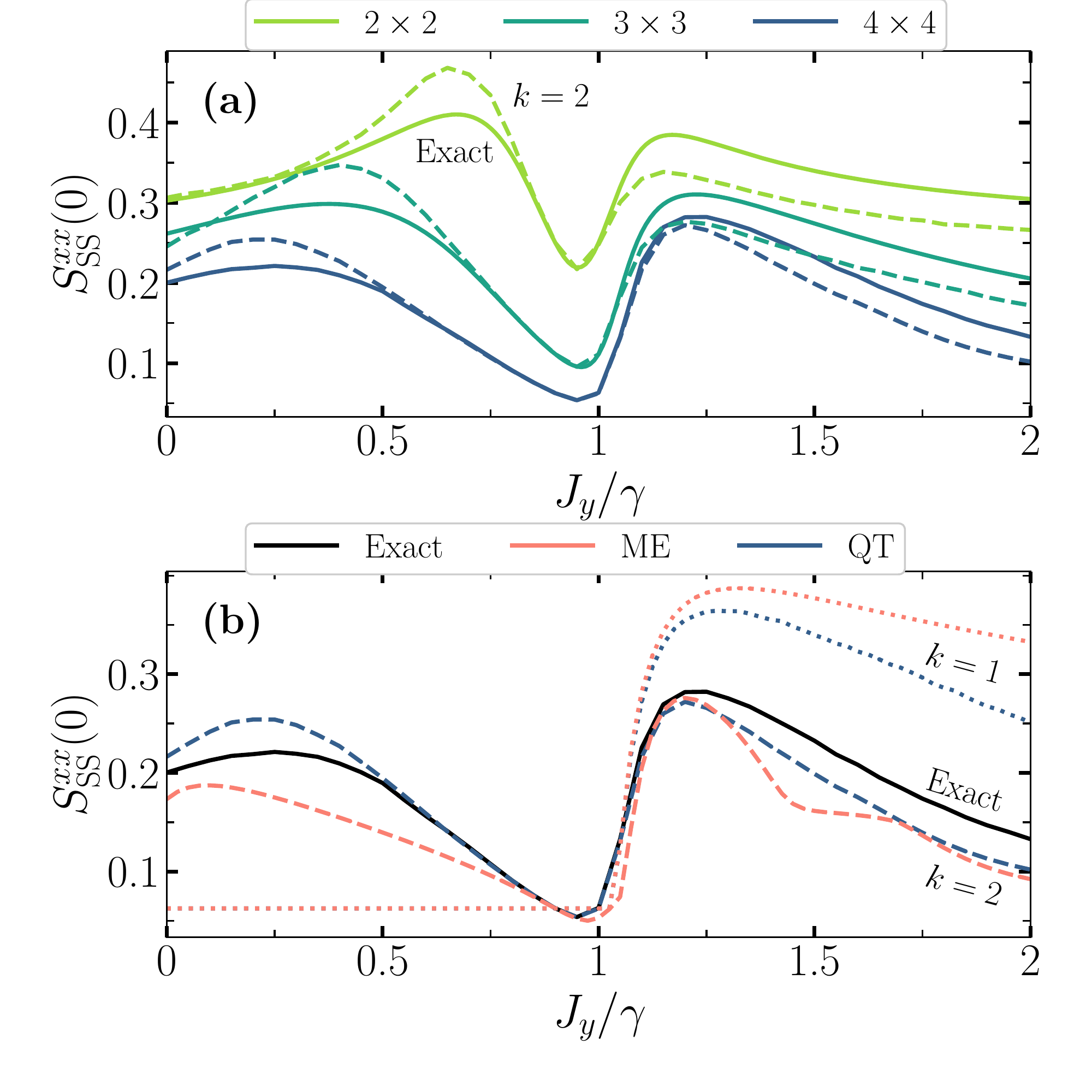}
    \caption{\textbf{(a)} Steady-state spin structure factor for the $\hat\sigma^x$ spin components, $S^{xx}_{\rm SS}(0)$,  for a $2\times2$, $3\times 3$ and $4\times 4$ lattice as a function of $J_y/\gamma$ ($J_x = 0.9 \gamma$): dashed lines represent the results from the $k=2$ truncation scheme, while the full lines are the exact steady state solution, obtained by directly solving the master equation exactly for $N < 4$, and by sampling exactly calculated trajectories in the stochastic unraveling of the master equation for $N = 4$. The parameter region in which the $k=2$ results overlap with the exact ones is seen to become progressively wider for larger system sizes. \textbf{(b)} Comparison of the same spin structure factor from different correlation-hierarchy methods, $k=1$ (dotted lines) and $k=2$ (dashed lines), for the $4\times 4$ lattice, applied at the level of the master equation (ME) and at the level of quantum trajectories (QT). For the data of both panels, and for all system sizes, the number of trajectories for the $k=2$ calculations is given by $N_{\rm traj} \approx 250$; the results are time-averaged over the time interval $t\gamma \in \left[75; 150\right]$, corresponding to the stationary regime of the evolution.}
    \label{fig:comp_exact}
\end{figure}

\subsection{The steady-state spin structure factor}\label{sec:structfact}
We start by benchmarking the $k=2$ truncation scheme, which amounts to tracking the evolution of one- and two-point correlators with ranges covering the entire lattice. To this aim, we compare our results on the steady-state expectation values with the exact solutions of a $2\times2$, $3\times 3$ and $4\times 4$ lattice. For the two smaller lattices we directly solve the master equation \eqref{eq:GKSL}; while the larger lattice is solved with a wave function Monte Carlo approach based on the photon counting unravelling of Eqs. \eqref{eq:pctrajeq}-\eqref{eq:pctrajeq2}, which converges to the exact steady-state expectation values when averaging over a sufficient number of trajectories.  The $\mathbb{Z}_2$ symmetry of the problem along the $x$ and $y$ axis  prevents the appearance of a nonzero magnetization in the steady state; hence we resort to the steady-state (SS) spin structure factor to reveal the possible appearance of ferromagnetic order in the system.  It is defined as
\begin{equation}\label{eq:steadystatespinstructurefactor}
    S_{\rm SS}^{\alpha\alpha}(\textbf{k}) = \frac{1}{N^2}\sum_{i,j} e^{i\textbf{k}.\textbf{r}_{ij}}\left\langle\hat{\sigma}_i^\alpha\hat{\sigma}_j^\alpha\right\rangle,
\end{equation}
with $\textbf{r}_{ij} = \textbf{r}_i - \textbf{r}_j$.
For the choice of $\textbf{k} = 0$, a value of the structure factor not scaling with the system size  signals the presence of long-range ferromagnetic order, while a value scaling to zero with system size (as $N^{-1}$) signals a paramagnetic phase. The results for the steady-state spin structure factor in the $x$-direction are shown in Fig. \ref{fig:comp_exact}(a) for values of $0 < J_y/\gamma < 2$. In the proximity of $J_y \approx \gamma$ we find excellent correspondence between our method and the exact solution. We refer the reader to Appendix~\ref{app:m2m4} for additional data showing a comparison of the second and fourth moments in the $x$-direction and $y$-direction for the $k=2$ trajectory method and the exact solution, once again exhibiting a rather remarkable correspondence.

Furthermore, as the system size is increased, so does the region where the $k=2$ results fall onto the exact solution. This can also be observed by studying the local maxima of the steady-state spin structure factor: for the maxima on both sides of $J_y = \gamma$ one can observe a decreasing difference with the exact solution as system size is increased, suggesting that a result very close to the exact solution may in fact be recovered in the thermodynamical limit.

The agreement between the numerically exact solution and the  $k=2$ results is especially good in the region close to the paramagnetic-to-ferromagnetic phase transition, occurring in the vicinity of $J_y/\gamma=1$ (as we shall see in Sec.~\ref{sec:phasetrans_expon}), and in a way which is nearly independent of system size. 
This good correspondence for small system sizes gives confidence for the use of our method to study the critical properties around the transition.
The very good agreement between the $k=2$  results and the exact ones suggests that the cumulants that we discard (of order $k=3$ and higher) are -as anticipated- significantly suppressed in the steady state as a result of the coupling to the environment. 

Fig.~\ref{fig:comp_exact}(b)) exposes the significant  improvement offered by the $k=2$ results compared to the $k=1$ ones throughout the range of parameters relevant for the physics discussed in this work.
The same figure shows as well that the application of the $k=2$ truncation scheme at the level of single-trajectory wavefunctions, which stochastically sample the density matrix, delivers results which are in significantly better agreement with the exact solution than those obtained by applying the same truncation scheme for the full fluctuations, i.e. at the level of the master equation. This is especially visible in the vicinity of the transition at $J_y/\gamma\approx 1$.

In the ferromagnetic phase, long-range ferromagnetic order appears in the $XY$-plane but not necessarily along one of the coordinate axes, and it is therefore necessary to systematically search for the direction of maximal correlations. Such a direction is defined by the angle $\phi$ for which the $k=0$ structure factor 
\begin{eqnarray}
    S^{\phi\phi}_{\rm SS}(0) & = & \cos^2 \phi  ~ S^{xx}_{\rm SS}(0) + \sin^2\phi ~S^{yy}_{\rm SS}(0) \nonumber \\
    && + 2 \sin\phi\cos\phi~ C^{xy}(0),
\end{eqnarray}
is maximal, where we have introduced the cross correlation term
\begin{equation}
C^{xy}(\textbf{k}) = \frac{1}{2N^2} \sum_{ij}e^{i\textbf{k}.\textbf{r}_{ij}} \langle \hat \sigma_i^x \hat \sigma_j^y + \hat \sigma_i^y \hat \sigma_j^x \rangle ~.
\end{equation}
Maximizing with respect to $\phi$, one can easily obtain the following condition on the optimal angle
 \begin{equation}\label{eq:optimalphi}
    \phi = \frac{1}{2}\tan^{-1}\left(\frac{2C^{xy}(0)}{S^{xx}_{\rm SS}(0) -S^{yy}_{\rm SS}(0)} \right) + \frac{p\pi}{2},
\end{equation}
where $p\in \mathbb{Z}$. This condition allows for the extraction of the angle $\phi$ which maximises the order parameter by simply calculating the three quantities $S^{xx}_{\rm SS}(0)$, $S^{yy}_{\rm SS}(0)$ and $C^{xy}(0)$. From this point onward we will systematically focus on results for the optimal angle $\phi$.  

\begin{figure}
    \centering
    \includegraphics[width=0.5\textwidth]{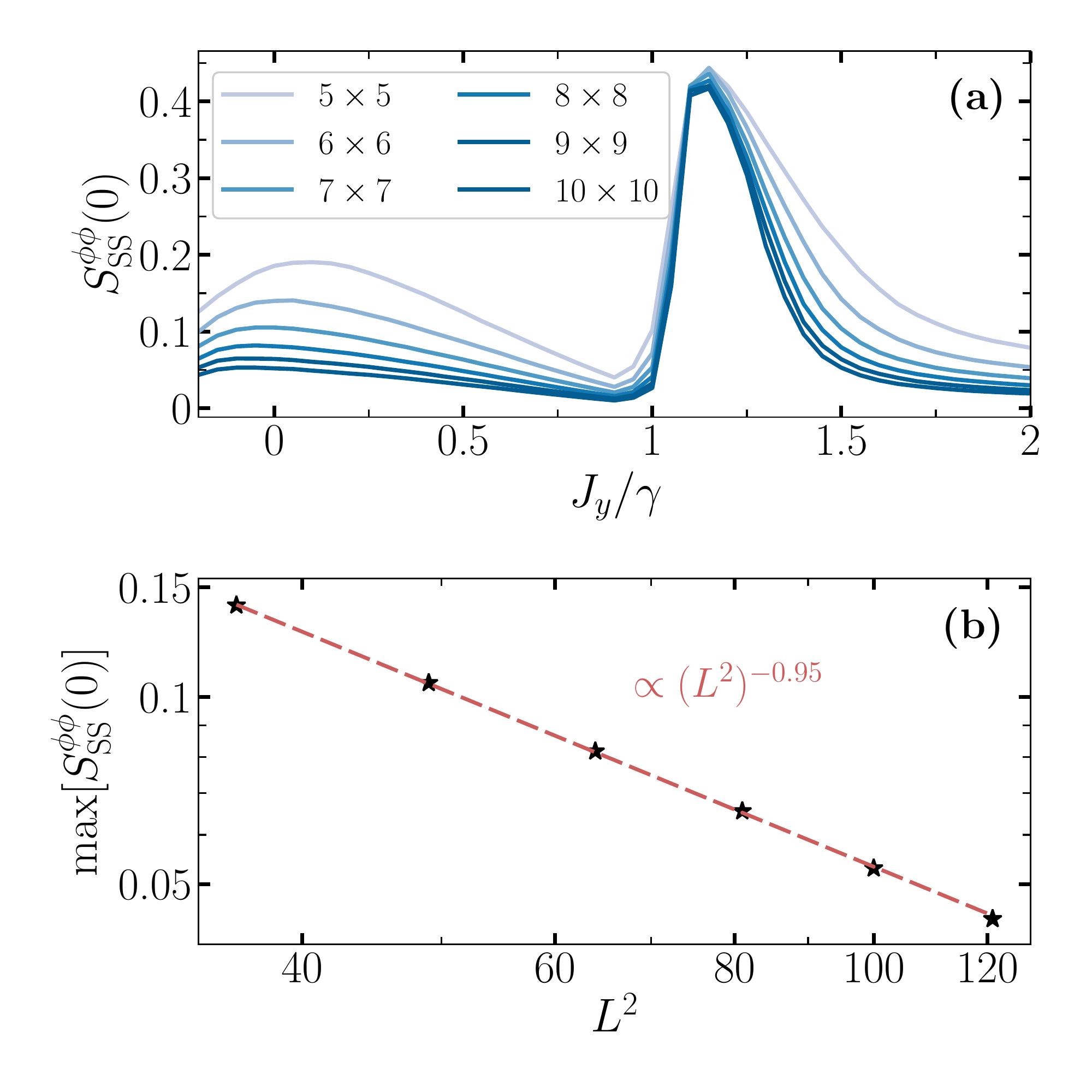}
    \caption{\textbf{(a)} Steady-state spin structure factor for the $\hat\sigma^\phi$ spin components at the optimal angle, $S^{\phi\phi}_{\rm SS}(0)$, for lattices with dimensions $L= 5, ..., 10$, obtained via quantum trajectories within the $k = 2$ truncation scheme ($J_x = 0.9 \gamma$). \textbf{(b)} Scaling of the local maximum of  $S^{\phi\phi}_{\rm SS}(0)$ for $J_y/\gamma < 0.9$;  the red line shows a power law fit $\propto (L^2)^{-0.95}$. In both panels, the other simulation parameters are the same as for the $k=2$ presented in Fig.~\ref{fig:comp_exact}.}
    \label{fig:ss_big}
\end{figure}

\subsection{Paramagnetic phase with quantum properties}
\label{sec:pm_quant}
The qualitative correspondence of the $k=2$ results allows us to investigate more closely the region $0< J_y/\gamma < 1$. In this region the Gutzwiller approach (either at the level of the master equation or of the wavefunction trajectories) predicts a featureless paramagnetic phase \cite{LeePRL13, CasteelsPRA18}, exhibiting a structure factor which is very close to zero for all system sizes.  On the other hand, our $k=2$ results show the existence of ferromagnetic correlations; the fact that they appear in this approach and not within the $k=1$ scheme \cite{HuybrechtsPRA19} indicates that they are have a quantum origin, given that the account of quantum correlations is the main distinction between the two truncation schemes. 
An analysis of scaling with system size allows us to determine the fate of these quantum ferromagnetic correlations in the thermodynamic limit.
In panel (a) of Fig. \ref{fig:ss_big} we show the steady-state spin structure factor $S_{SS}^{\phi\phi}$ for various system sizes. It is clear that the maximum for the left peak decreases with increasing system sizes, but correlations persist for lattices of intermediate size. A finite-size scaling of the maximum, shown in panel (b) of Fig. \ref{fig:ss_big}, shows a power law decrease proportional to $N^{-0.95}$. This signals that in fact the quantum ferromagnetic correlations are of short-range nature, as we will show explicitly in the next section, and one recovers a paramagnetic phase as predicted by the Gutzwiller trajectory approach \cite{HuybrechtsPRA19}.
Nonetheless, as shown in Sec.~\ref{sec:squeezing}, the quantum nature of correlations present in this phase is not only underlined by the comparison with the Gutzwiller results, but it can be further characterized in terms of entanglement witnesses. Indeed, as we shall see, substantial quantum correlations are accompanied by spin squeezing, offering a rather tight lower bound to the quantum Fisher information associated with the collective spin in some parameter regimes.

\subsection{Long-range order and short-range quantum correlations}
\label{sec:corrfunc}

\begin{figure}
    \centering
    \includegraphics[width=0.5\textwidth]{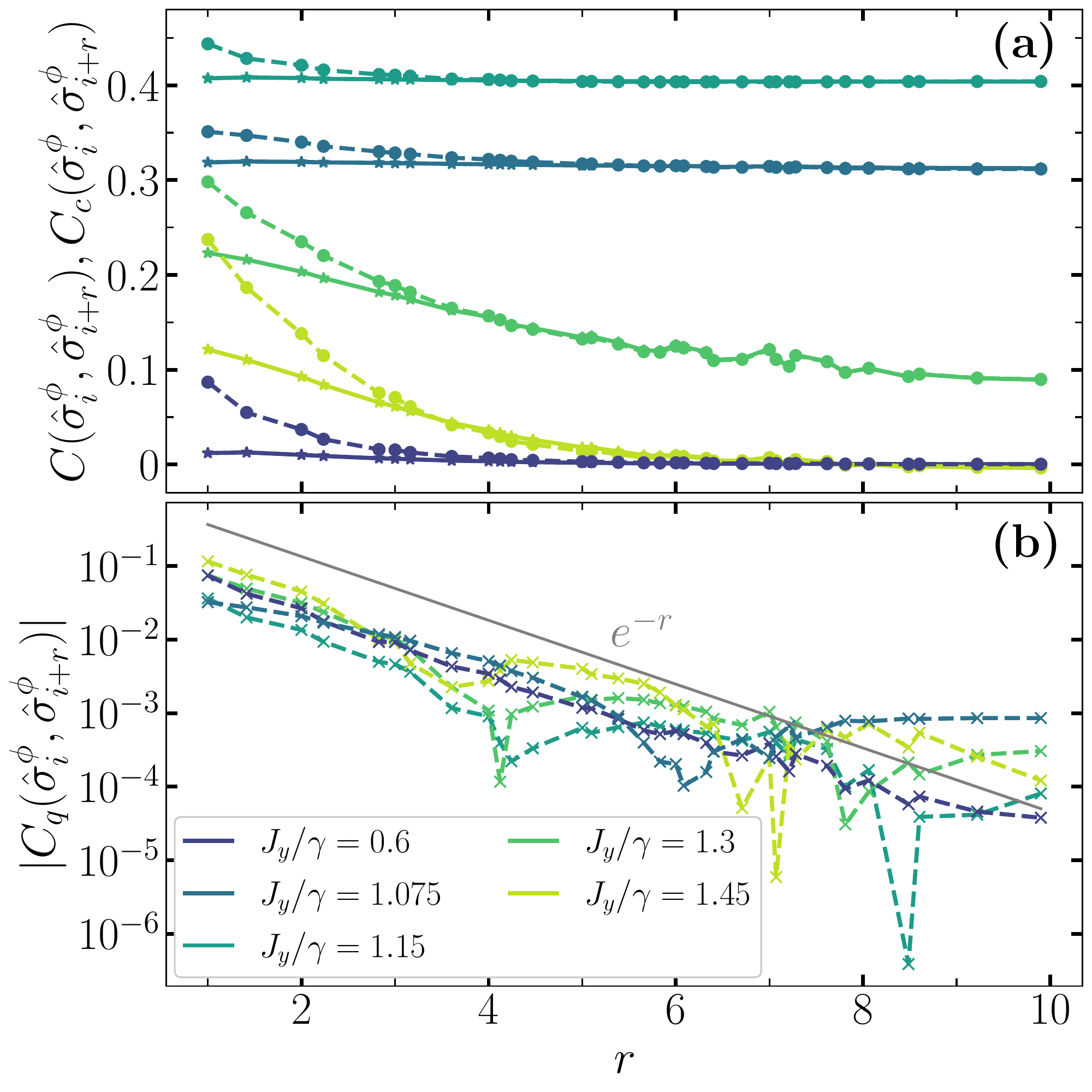}
    \caption{\textbf{(a)}) Total correlation function $C(\hat\sigma_i^\phi, \hat\sigma_{i+r}^\phi)  = \langle \hat\sigma_i^\phi \hat\sigma_{i+r}^\phi \rangle$ for the optimal-angle spin components (circle-marked dashed lines), along with its classical part $C_c(\hat\sigma_i^\phi, \hat\sigma_{i+r}^\phi)$ (star marked full lines), for several values of $J_y/\gamma$ at $J_x = 0.9 \gamma$. \textbf{(b)} Quantum contribution to the correlations $C_q(\hat\sigma_i^\phi, \hat\sigma_{i+r}^\phi)$ (cross-marked dashed lines), compared with a reference exponential decay (full line). 
  In both panels, the data refer to a $14\times 14$ lattice,  and were obtained using $N_{\rm traj} = 128$ trajectories, with time-averaging performed in the time interval $t\gamma \in \left[75; 150\right]$. 
  }
    \label{fig:corrfunction14}
\end{figure}

The most salient feature in Fig. \ref{fig:ss_big}(a) is the appearance of a strong peak in the structure factor, which is nearly size independent in the region 
$1 \lesssim J_y/\gamma \lesssim 1.25$, reflecting the appearance of long-range ferromagnetic order. The phase diagram therefore sees the succession of two transitions upon increasing $J_y/\gamma$, from paramagnetic to ferromagnetic around $J_y/\gamma \approx1$; and from ferromagnetic back to paramagnetic around $J_y/\gamma \approx 1.25$. The critical behavior at these two transitions will be investigated in details in the following Sec.~\ref{sec:phasetrans_expon}.

We shall now examine the steady-state correlation function $\left\langle \hat{\sigma}_i^\phi\hat{\sigma}_j^\phi\right\rangle$, for the optimal angle $\phi$ spin components, i.e. $\hat{\sigma}^\phi_i = 
\cos(\phi)\hat{\sigma}_i^x + \sin(\phi)\hat{\sigma}_i^y$. These correlations are shown in Fig.~\ref{fig:corrfunction14} (a) for a $14\times14$ lattice, for various values of $J_y/\gamma$ by the dashed lines with circle markers. In this figure we plot, along with the above cited correlation function, the one associated to classical correlations, as defined in Eq.~\eqref{eq:classcorr}, namely accounting only for trajectory-to-trajectory fluctuations of the single-trajectory average values $\langle \psi_n | \hat{\sigma}_i^\phi |\psi_n \rangle$. The difference between these data give in turn the quantum correlations as defined in Eq.~\eqref{eq:qcorr}.

We systematically observe for all values of $J_y/\gamma$ that classical correlations and total correlations tend to coincide at long distances, indicating that the paramagnetic-to-ferromagnetic and ferromagnetic-to-paramagnetic transitions in the system are fundamentally driven by classical fluctuations -- this conclusion will also be corroborated in the following Sec.~\ref{sec:phasetrans_expon} by the analysis of the universality class of the transition. Nonetheless we clearly observe the presence of very pronounced \emph{short-range} quantum correlations, in that classical and total correlations significantly deviate from each other at shorter distances. More precisely, as shown in Fig.~\ref{fig:corrfunction14} (b), we observe exponentially decaying quantum correlations exhibiting a finite quantum coherence length $\tilde{\ell}_Q$ as defined in  Sec.~\ref{sec:bounds}. Remarkably, we observe $\tilde{\ell}_Q\approx 1$, the lattice constant, across all parameter values.
The ability of the method to tackle rather large lattices allows us to describe the full spatial structure of quantum correlations. Their short-ranged nature may erroneously suggest that one may ignore them altogether (as done by Gutzwiller trajectory approaches, or their cluster extensions for distances beyond the cluster size); and that this will not bear any consequence on the study of the critical behavior of the system -- which by definition only involves long-range correlations. 
In fact we shall see in the next section that taking quantum correlations into account properly has profound consequences for the critical behavior, in spite of its seemingly classical nature. 

\begin{figure}
    \centering
    \includegraphics[width=0.5\textwidth]{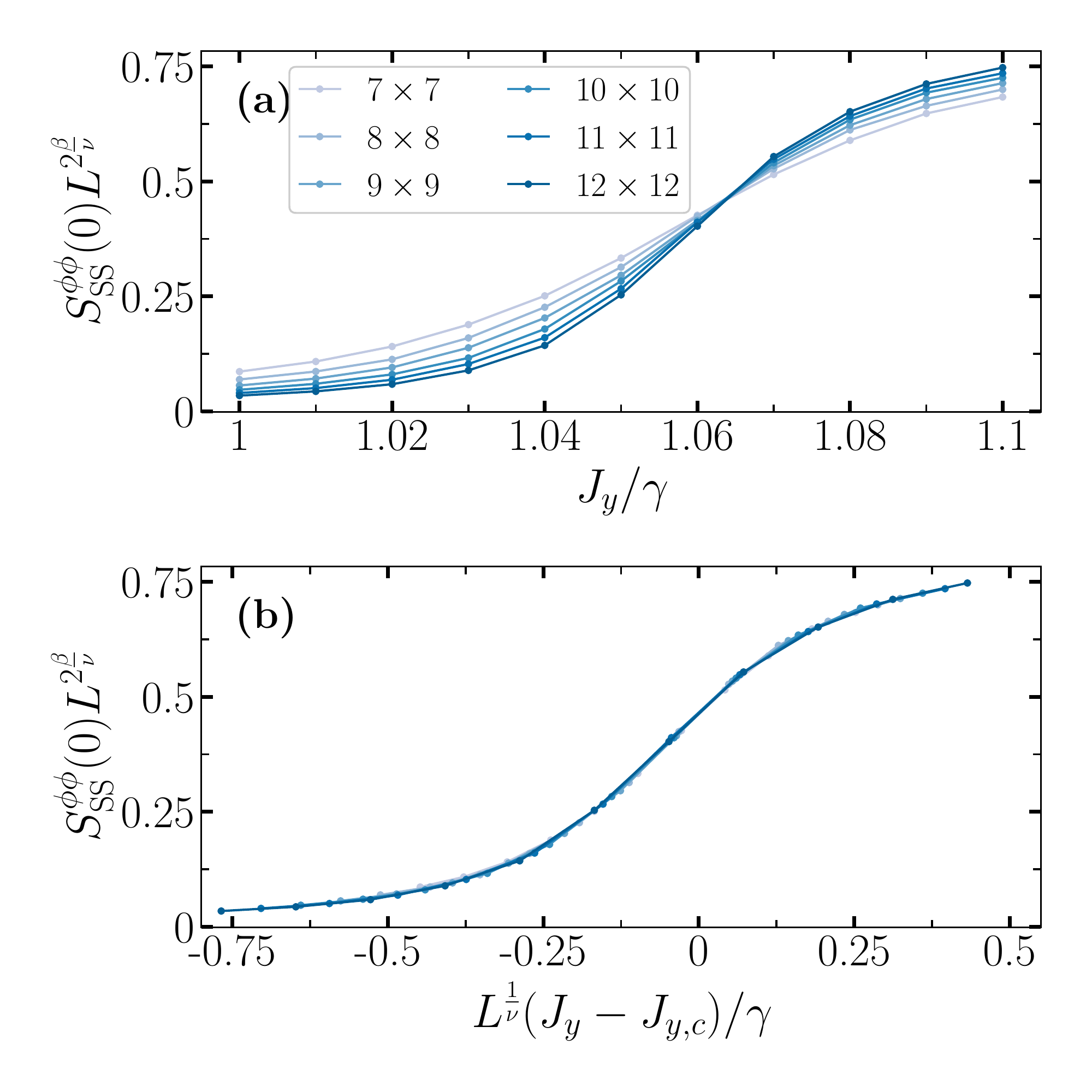}
    \caption{\textbf{(a)} Rescaled structure factor $S_{\rm SS}^{\phi\phi}(0) L^{2\beta/\nu}$ at the paramagnetic-to-ferromagnetic transition for $J_x = 0.9 \gamma$, using 2D Ising exponents ($\nu = 1$ and $\beta = \frac{1}{8}$); \textbf{(b)} Full scaling plot, using $J_{y,c} \approx 1.064$ as critical point. For all system sizes the number of trajectories is given by $N_{\rm traj} \approx 768$, and time-averaging is performed in the interval $t\gamma \in \left[75; 150\right]$.}
    \label{fig:exponentsleft}
\end{figure}

\begin{figure}
    \centering
    \includegraphics[width=0.5\textwidth]{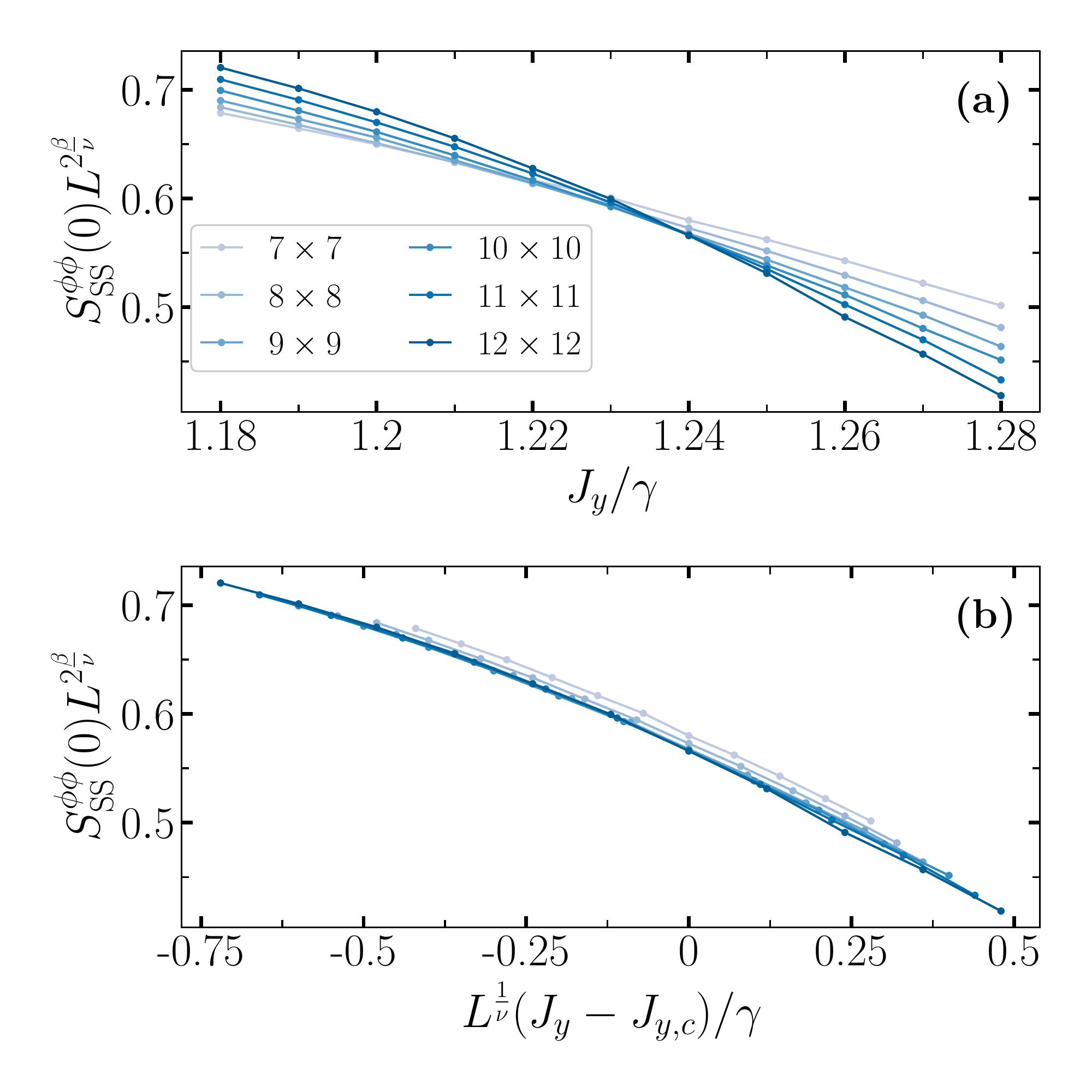}
    \caption{\textbf{(a)} Rescaled structure factor $S_{\rm SS}^{\phi\phi}(0) L^{2\beta/\nu}$ at the ferromagnetic-to-paramagnetic transition for $J_x = 0.9 \gamma$, using 2D Ising exponents ($\nu = 1$ and $\beta = \frac{1}{8}$). Inset: crossing points $J_{y,c}$ of the curves associated with sizes $L$ and $L+1$ of panel (a); \textbf{(b)} Full scaling plot, using $J_{y,c} \approx 1.24$ as critical point. For all system sizes the number of trajectories is given by $N_{\rm traj} \approx 1408$ for $L = 7,8,9$, $N_{\rm traj} \approx 1920$ for $L = 10, 11$, and $N_{\rm traj} \approx 1792$ for $L = 12$; and time-averaging is performed in the interval $t\gamma \in \left[75; 150\right]$.}
    \label{fig:exponentsright}
\end{figure}

\subsection{Phase transitions and universal behavior}\label{sec:phasetrans_expon}
We now turn to a systematic finite-size scaling analysis of the two transitions appearing in the system: the paramagnetic-to-ferromagnetic transition for $J_y/\gamma \approx 1$, and the transition to re-entrant paramagnetism for $J_y/\gamma \approx 1.25$. While the presence of the first phase transition is well established, the second one is debated, and even proposed to be in fact a smooth crossover \cite{RotaNJP18}. In particular a difficult aspect for this transition is that it occurs in a regime in which the steady state has high entropy, posing a challenge to all density-matrix methods which are limited in the entropy content of the state \cite{FinazziPRL15, RotaPRB17}. 
We argue that \emph{a priori}, our method should be able to capture the proper critical behavior regardless of whether it is driven by classical or quantum fluctuations.

\begin{figure*}
    \centering
    \includegraphics[width=0.8\textwidth]{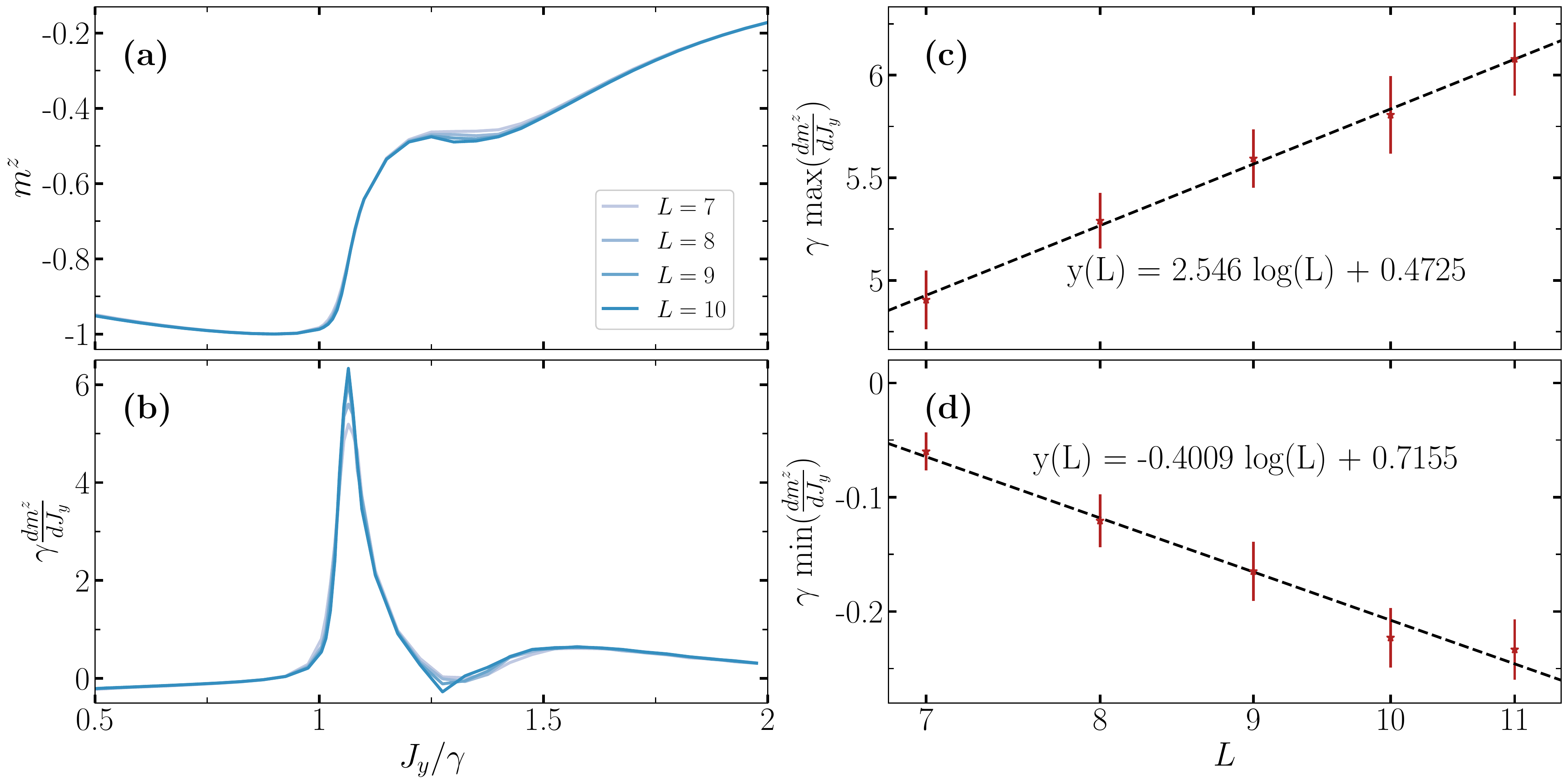}
    \caption{\textbf{(a)} Dissipation-induced transverse magnetization $m^z$  as a function of $J_y$ for $J_x = 0.9 \gamma$ and different system sizes; \textbf{(b)} Derivative of the transverse magnetization $\gamma \dd{m^z}/\dd{J_y}$, showing two clear anomalies at the two transitions of the system; \textbf{(c)} scaling of the peak value of $\gamma \dd{m^z}/\dd{J_y}$ with system size (around $J_y/\gamma \approx 1$); \textbf{(d)} scaling of $\gamma \dd{m^z}/\dd{J_y}$ at the minimum value (around $J_y/\gamma \approx 1.25$). In both panels (c) and (d) the dashed line is a logarithmic fit to the data. The number of trajectories used in (a) and (b) are $N_{\rm traj}\approx 250$ and $N_{\rm traj}\approx 750$ in the region close to $J_y 
    \approx 1.05$. In panel (c) the number of trajectories range from $N_{\rm traj}\approx 27000$ to $N_{\rm traj}\approx 15000$ and in panel (d)  from $N_{\rm traj}\approx 70000$ to $N_{\rm traj}\approx 35000$ (depending on lattice dimension). For each trajectory time-averaging is performed in the interval $t\gamma \in \left[75; 150\right]$.}
    \label{fig:magderivative}
\end{figure*}

\subsubsection{Structure factor} 

In analogy with equilibrium thermal phase transitions, we shall assume that the behavior of the system exhibits scale invariance at a dissipative phase transition, so that the singular part of all thermodynamic quantities exhibits in turn a scaling behavior, governed by critical exponents. As a consequence of such scaling behavior, the structure factor at the optimal angle $\phi$ is expected to exhibit the following scaling behavior on finite-size systems: 
\begin{equation}
    S_{\rm SS}^{\phi\phi}(0)=L^{-2\beta/\nu}F(\vert J_y-J_{y,c}\vert L^{1/\nu}),
\end{equation}
where $J_{y,c}$ is the critical value for our control parameter $J_y$; $L$ is the linear system size; $F$ is a universal scaling function; and $\beta, \nu$ are universal critical exponents. 
In order order to extract the three parameters  $J_{y,c}$,  $\beta$ and $\nu$ from a finite-size-scaling analysis of our $k=2$ results, we should adjust the values of the parameters so that the curves for $S_{\rm SS}^{\phi\phi}(0) L^{2\beta/\nu}$ plotted as a function of $\vert J_y-J_{y,c}\vert L^{1/\nu}$ for different system sizes collapse together, reconstructing the universal scaling function $F$. Reducing the number of fitting parameters, we start off with an educated guess; and immediately observe the consistence of our results with the 2D Ising universality class. Indeed the insight gained by the analysis of correlations  in the previous section showed us that long-range correlations in the system are of classical origin, so that we should expect the transitions (from paramagnetic to ferromagnetic and back) to be of classical nature, and compatible with the $\mathbb{Z}_2$ symmetry of the system, its two-dimensional nature, and the short-range nature of its couplings. These characteristics suggest the 2D classical Ising universality class as a natural candidate for our transition, inviting us to fix the critical exponents to the corresponding values $\beta = 1/8$ and $\nu=1$. Therefore the only parameter to be adjusted remains $J_{y,c}$.

The value of $J_{y,c}$  can be extracted as the crossing point $J_y$ between curves of the rescaled structure factor $S_{\rm SS}^{\phi\phi}(0) L^{2\beta/\nu}$ for different system sizes.  Our results show a crossing at $J_y/\gamma \approx 1.064$ (Fig.~\ref{fig:exponentsleft}(a)) for the first transition, and at $J_y/\gamma \approx 1.24$ (Fig.~\ref{fig:exponentsright}(a)) for the second one. The proof that the transition belongs to the classical 2D Ising universality class comes when plotting the rescaled structure factor as a function of the rescaled distance to the critical point $\vert J_y-J_{y,c}\vert L^{1/\nu}$ with $\nu = 1$: Figs.~\ref{fig:exponentsleft}(b) and \ref{fig:exponentsright}(b) show an excellent collapse for the first transition, and a very good one for the second transition when looking at the largest lattices (from $9\times 9$ to $12 \times 12$). The apparently imperfect collapse at the second transition for smaller system sizes is clearly due to finite-size effects, as already visible in Fig.~\ref{fig:exponentsright}(a), in which the crossing point of the curves stabilizes only starting from the $9 \times 9$ lattice, as can be seen in the inset of Fig.~\ref{fig:exponentsright}(a). The above results provide therefore conclusive evidence for the existence of two transitions, belonging both to the 2D classical Ising universality class. 

As discussed in Appendix \ref{ap:k1}, a similar scaling analysis shows that the results obtained within the $k=1$ (Gutzwiller) scheme are \emph{not} compatible with a 2D Ising transition, nor with a mean-field one -- in fact we can obtain a scaling collapse of our $k=1$ data only for effective critical exponents that do not correspond to any equilibrium universality class known to us. This result is rather surprising, in view of the fact that one would expect the $k=1$ approach to capture classical fluctuations at the dissipative transition, and that such fluctuations are expected to govern the critical behavior. From this observation we conclude that short-range quantum correlations, included in the $k=2$ approach, are essential in determining the universality class, even though the long-range correlations that emerge at criticality are of classical origin. We can partially attribute this to the fact that the diffusion constants in the $k=1$ model become zero in the paramagnetic phase so that no noise is left there. This aspect is rather surprising, but it shows that dissipative phase transitions often defy the intuition for critical phenomena that one may have developed in the context of equilibrium systems. Indeed we can put this observation in parallel with the (equally surprising) one that cluster mean-field approaches \cite{JinPRX16}, only including short-range correlations, change radically the prediction for the phase diagram of our system of interest compared with the standard mean-field approach  \cite{LeePRL13}. In both cases, one observes that the proper account of fluctuations at short scales in dissipative quantum systems can have significant consequences on the long-wavelength properties.

\subsubsection{Derivative of the transverse magnetization}

The steady state of the dissipative XYZ model at study is generally characterized by the presence of a net magnetization $m^z = \langle \hat J^z \rangle /N = N^{-1} \sum_i \langle \hat \sigma_i^z \rangle$ along the $z$ axis -- which is induced by the fact that dissipation in the form of spontaneous decay favors the spin to point downwards along this axis. The $m^z$ magnetization takes value of $-1$ at the U(1) symmetry point of the model ($J_x = J_y$), at which the steady state is fully polarized along $-z$ by the dissipation given that $M^z$ is a good quantum number. But it decreases (in absolute value) with respect to is saturation value as soon as $J_y \neq J_x$ because of quantum effects, given that the Hamiltonian $\hat H$ ceases to commute with $\hat J^z$. Upon increasing the value of $J^y/\gamma$ at fixed $J^x/\gamma$, the $m^z$ curve exhibits clearly two size-dependent anomalies, corresponding to the two transitions of the system: as shown in Fig.~\ref{fig:magderivative}(a): a sharp decrease (in absolute value) at the first (paramagnetic-to-ferromagnetic) transition; and a successive upturn (again in absolute value) at the second (ferromagnetic-to-paramagnetic) transition. These features are best captured by taking the derivative of the magnetization with respect to the control parameter of the transition $\gamma \dd{m^z}/\dd{J_y}$: this derivative -- shown in Fig.~\ref{fig:magderivative}(b) -- exhibits two sharp size-dependent features, namely a sharp growing peak and an equally sharp growing dip. Tracking the size dependence of the height of the peak (at the paramagnetic-to-ferromagnetic transition) we observe that it is compatible with a logarithmic growth $\gamma \left ( \dv{m^z}{J_y} \right)_{\rm peak} \approx A + B \log L$ (Fig.~\ref{fig:magderivative}(c)). A similar behavior is observed as well for the dip in the derivative at the second transition, as shown in Fig.~\ref{fig:magderivative}(d). A logarithmic growth is to be expected according to the 2D Ising universality class: indeed a similar logarithmic divergence of the derivative of the transverse magnetization with respect to the control parameter of the transition (the temperature, in this case) is observed at the thermal transition of the 2D Ising model in a transverse field, and, as discussed in Appendix~\ref{ap:z-mag} from the scaling form of the free energy, it can be proven to be equivalent to the well-known logarithmic divergence of the specific heat peak at the 2D Ising transition \cite{kardar2007statistical}. Therefore this result corroborates further the adherence of the two transitions of the system to the 2D Ising universality class; as well as the ability of the $k=2$ truncation scheme approach to dissipative phase transitions to reconstruct accurately the multiple facets of critical behavior. 

\subsection{Bounds on the Quantum Fisher Information}\label{sec:squeezing}

\subsubsection{Spin squeezing as an entanglement witnesses}

In Sec.~\ref{sec:corrfunc} we have already ascertained the existence of short-range quantum correlations along each stochastic trajectory. Nonetheless, this result is strongly dependent on the properties of the trajectory wavefunctions, and it could in principle be interpreted as depending on the specific unraveling that we are considering.  Nonetheless, as already discussed in Sec.~\ref{sec:bounds} the integral of the (unraveling-dependent) quantum correlations 
\begin{equation}
\begin{split}
    F_q(\hat{J}^\phi) &= \sum_{i,j} \Big[\cos^2 \phi ~ C_q(\hat \sigma_i^x, \hat \sigma_j^x) + \sin^2 \phi ~ C_q(\hat \sigma_i^y, \hat \sigma_j^y) \\
    &+ \frac{1}{2}\sin(2\phi) \left( C_q(\hat \sigma_i^x, \hat \sigma_j^y) + C_q(\hat \sigma_i^y, \hat \sigma_j^x) \right ) \Big ]~.
    \end{split}
\end{equation}
provides an upper bound to the QFI of the collective spin component along the optimal angle $\hat J^\phi = \sum_i  \sigma^\phi_i$, ${\rm QFI}(\hat J^\phi)$; and the spatial decay of quantum correlations defines a similar bound to the spatial decay of the QFIM. Here we shall discuss how our calculations can access in turn a \emph{lower} bound to  ${\rm QFI}(\hat J^\phi)$, allowing therefore for a quantitative estimate of its value. 

The quantity of interest is related to the \emph{spin squeezing parameter}, which probes the structure of the uncertainty on the orientation of the collective spin $\hat J^\alpha = \sum_i  \sigma^\alpha_i$ $(\alpha = x,y,z)$. 
We observe that the state of the system is magnetically polarized along the negative $z$ direction for the spins because of spontaneous decay -- namely it develops a finite value for $\langle J^z\rangle$ in the steady state.  Moreover at the transition the uncertainty on the collective spin component at the optimal angle $\phi$, $\hat J^{\phi} = \cos\phi ~ \hat J^x +\sin\phi ~ \hat J^y $, develops anomalous critical fluctuations; if these fluctuations have an enhanced quantum component, then one can expect that anomalously small fluctuations are developed by the perpendicular collective spin component $J^{\phi_\perp}$, as observed at Ising quantum critical points \cite{FrerotR2018}. Under these circumstances, entanglement can be effectively detected in the form of squeezing, namely by the fact that the squeezing parameter \cite{WinelandPRA1994} 
\begin{equation}
    \xi_R^2 = \frac{N~ \text{Var}\big(\hat{J}^{\phi_\perp}\big)}{\langle\hat{J}^z\rangle^2},
    \label{eq:xi2R}
\end{equation}
becomes smaller than unity, or equivalently $\xi_R^{-2} > 1$. This condition is enough to show that the state is not separable \cite{Sorensen2001}. 

The inverse of the squeezing parameter inherits its entanglement witnessing properties (discussed in Sec.~\ref{sec:bounds}) from the fact of being a lower bound to the quantum Fisher information (QFI) associated with the most strongly fluctuating collective spin component, ${\rm QFI}(\hat J^{\phi})$. 
Indeed the inverse squeezing parameter represents a \emph{lower} bound to the QFI density, $\xi_R^{-2} \leq {\rm QFI}(\hat J^{\phi})/N$ \cite{PezzeRMP2018}, as it offers the gain in metrological precision (compared to the SQL) using a specific measurement protocol (Ramsey interferometry \cite{WinelandPRA1994}). Therefore the entanglement witnessing properties of the QFI are directly transferred to the inverse spin-squeezing parameter.

\subsubsection{Results} 

\begin{figure}
    \centering
    \includegraphics[width=0.5\textwidth]{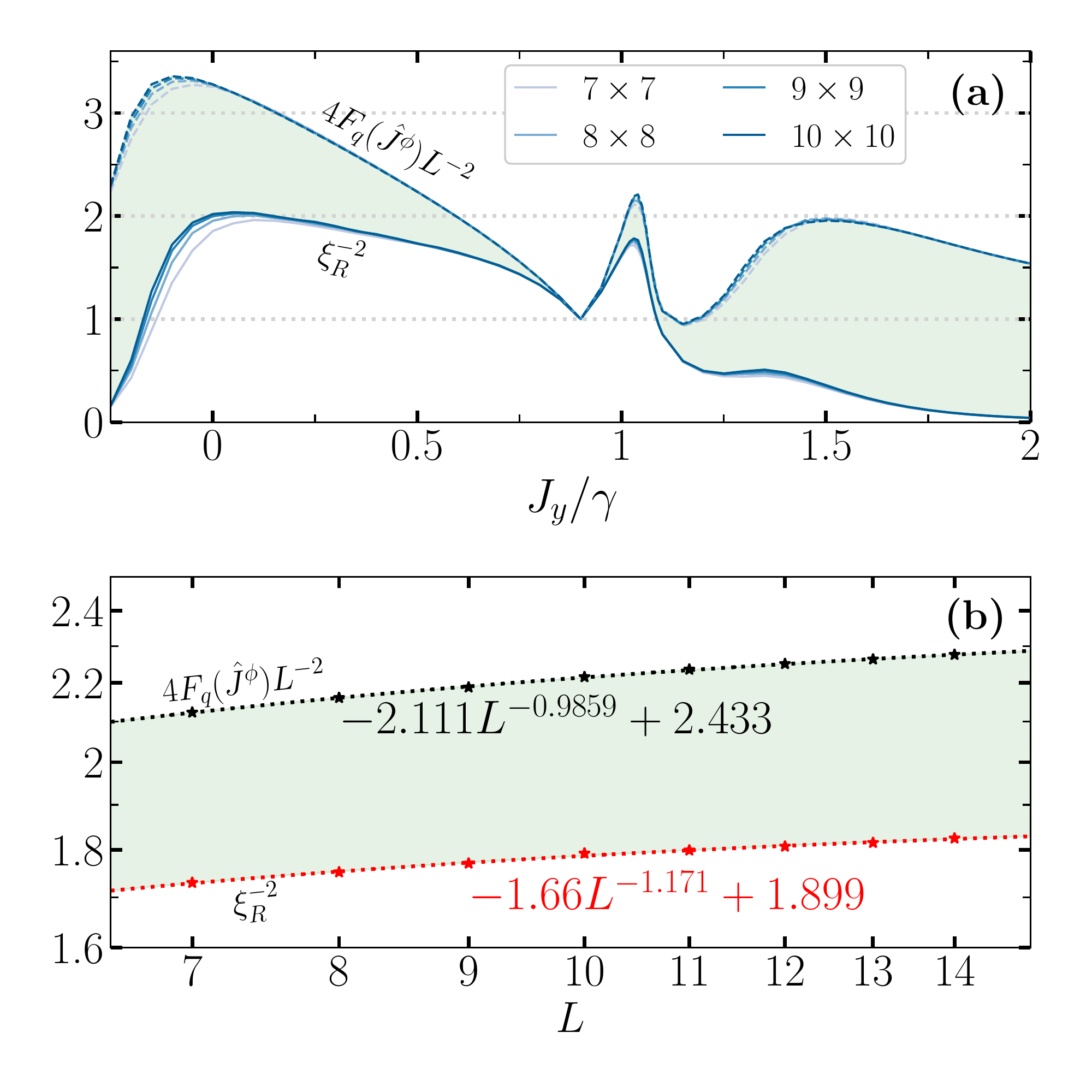}
    \caption{\textbf{(a)} Inverse spin-squeezing parameter $\xi_{R}^{-2}$ (lower solid curves) and $4F_q(\hat J^{\phi})$ (upper dashed curves), as a function of $J_y/\gamma$ ($J_x/\gamma = 0.9$) for different system sizes. The two sets of curves provide a lower and upper bound to $\text{QFI}(\hat{J}^\phi)$ respectively (see Eq.~\eqref{eq:chain}), whose value is therefore comprised within the green-shaded area (for a $10\times 10$ lattice).
    For all system sizes the number of trajectories is given by $N_{\rm traj} \approx 770$, and time-averaging is performed in the time interval $t\gamma \in \left[75; 150\right]$. \textbf{(b)} Scaling of the maximum of  $\xi_R^{-2}$ and of $4F_q(\hat J^{\phi})$ near the paramagnetic-to-ferromagnetic transition ($J_y\approx 1.064\gamma$).}
    \label{fig:bounds}
\end{figure}

Our estimate of the QFI of the steady state proceeds then by exploiting the inequality chain
\begin{equation}
\xi_R^{-2} \leq \frac{{\rm QFI}(\hat J^{\phi})}{N}  \leq \frac{4 F_q(\hat J^{\phi})}{N} ~
\label{eq:chain}
\end{equation}
and by calculating explicitly the two bounds. Fig.~\ref{fig:bounds} (a) shows the inverse squeezing parameter (full lines) for different system sizes as a function of $J_y/\gamma$ scanning across the two dissipative phase transitions of the system. We observe that squeezing (namely the condition $\xi_R^{-2} >1$) is massively present in the phase diagram of the system: in particular squeezing accompanies the first paramagnetic-to-ferromagnetic transition at $J_y/\gamma \approx 1$; and, most prominently, it is present across a wide region of the paramagnetic phase for $J_y/\gamma < 1$, which, as already pointed out in Sec.~\ref{sec:pm_quant}, is accompanied by pronounced quantum correlations. The cusp singularity of squeezing for $J_y/\gamma = J_x/\gamma = 0.9$  marks the fact that the steady state at this symmetry point is the factorized pure state $|\downarrow \downarrow ... \downarrow\rangle$. The fact that squeezing is nearly independent of system size is a reflection of the fact that quantum correlations are short-ranged, as already explicitly shown in Sec.~\ref{sec:corrfunc}. 
At the same time, squeezing is absent at the second transition for $J_y/\gamma \approx 1.2$. The striking difference between the behavior at small $J_y/\gamma$ vs. the behavior at larger $J_y/\gamma$ is certainly a consequence of the fact that the regime at larger $J_y/\gamma$ exhibits much larger entropies \cite{RotaPRB17}, so that quantum-coherence effects are expected to be suppressed. 

While the presence of squeezing is conclusive proof for the entangled nature of the state, its absence does not allow one to draw any conclusion on the nature of the state, since entanglement may still be witnessed by another criterion. Such a criterion could be offered by the QFI density exceeding unity, which is more effective than the squeezing criterion as, by construction, it detects all metrologically useful forms of entanglement, irrespective of the measurement protocol used to exploit it. 
In Fig.~\ref{fig:bounds} (a) we show both bounds for the QFI density appearing in Eq.~\eqref{eq:chain}, as they evolve across the two phase transitions. There we observe that the inequality chain becomes tight in the vicinity of the factorization point $J_y/\gamma = 0.9$ and of the first transition, revealing that squeezing is in fact the nearly optimal metrological resource of the steady state of the system, ensuring a sensitivity of the state to rotations which exceeds the SQL. On the other hand, the bound becomes looser for smaller values of $J_y/\gamma$ as well as larger ones. This may mean that the heterodyne unravelling is simply far from the one minimizing the quantum fluctuations on pure state decompositions in Eq.~\eqref{eq:upperbound}; or that metrologically useful entanglement exists in these regimes instead, but in forms different from (or superior to) spin squeezing. The conclusive aspect of our analysis in these regimes is that the upper bound on the QFI density, $4 F_q(\hat J^{\phi})$, does \emph{not} appear to scale with system size, which implies that the QFI density itself cannot scale either: this is yet another consequence of the short-range nature of quantum correlations, pointed out in Sec.~\ref{sec:corrfunc}. To corroborate this observation, an analysis of the scaling near the paramagnetic to ferromagnetic transition where $J_y\approx\gamma$, is shown in Fig.~\ref{fig:bounds} (b). Both the lower bound and upper bound on the QFI density converge to a finite value in the thermodynamic limit, resp. $\approx 1.9$ and $\approx 2.4$ as indicated by their respective fits. Hence the QFI density, albeit not scalable, is predicted to witness entanglement in the infinite-size limit.

\begin{figure}
    \centering
    \includegraphics[width=0.5\textwidth]{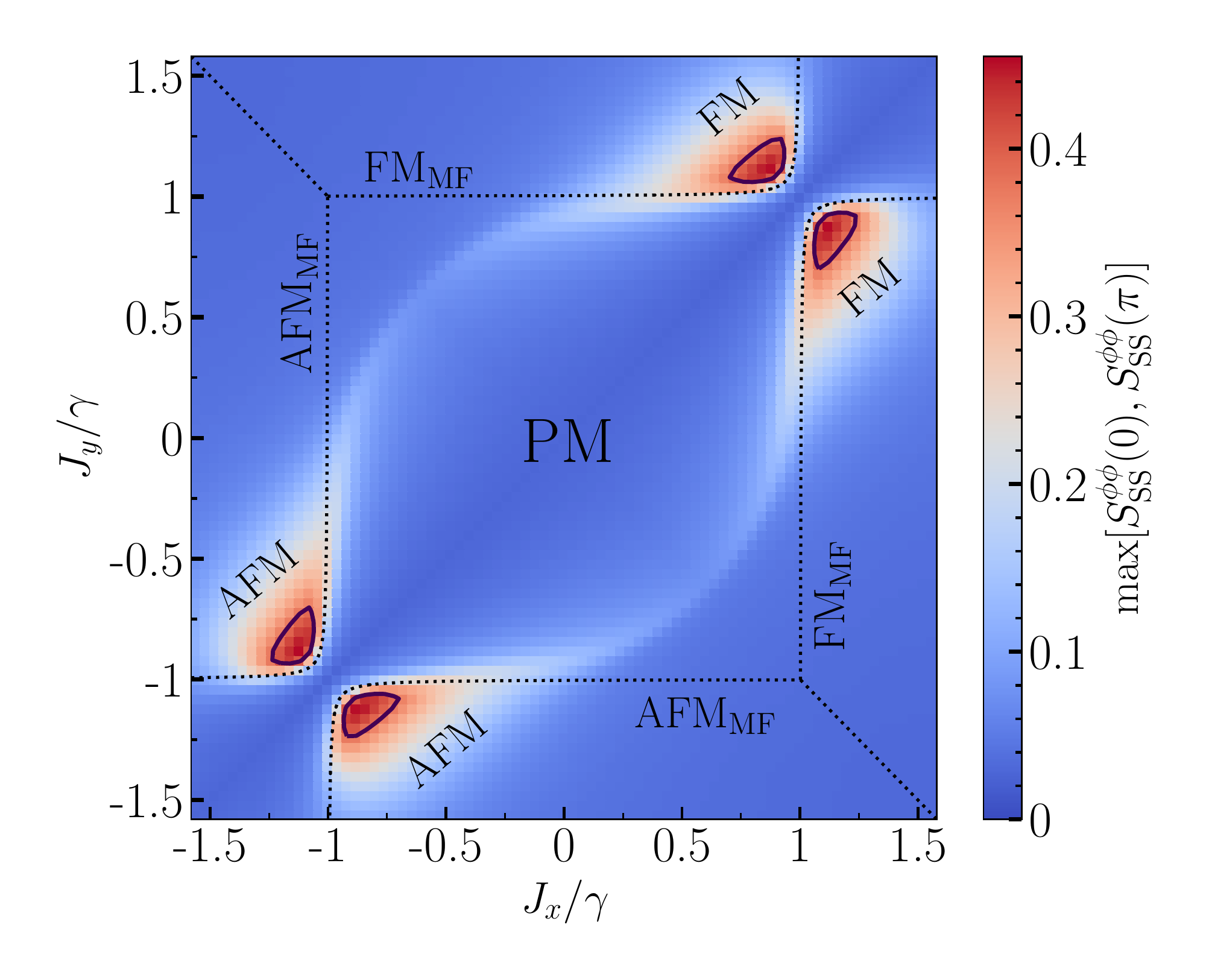}
    \caption{Maximum of the structure factor, $\max[S_{\rm SS}^{\phi\phi}(0), S_{\rm SS}^{\phi\phi}(\pi,\pi)]$ for a $6\times 6$ lattice with $J_z = \gamma$. The black dotted lines show the mean-field prediction for the boundaries of the FM and AFM phase. The full black lines show the contour line where the (maximal) structure factor is equal to its value at the critical point $(J_x, J_{y,c})/\gamma = (0.9, 1.24)/\gamma$, repeated three times by reflection symmetry around the $J_x = J_y$ axis and around the $J_x = -J_y$ axis.
    Each data point is obtained with a number of trajectories $N_{\rm traj} \leq 320$, and time-averaging is performed in the time interval $t\gamma \in [50,150]$. }
    \label{fig:m2phiN7}
\end{figure}

\begin{figure}
    \centering
    \includegraphics[width=0.5\textwidth]{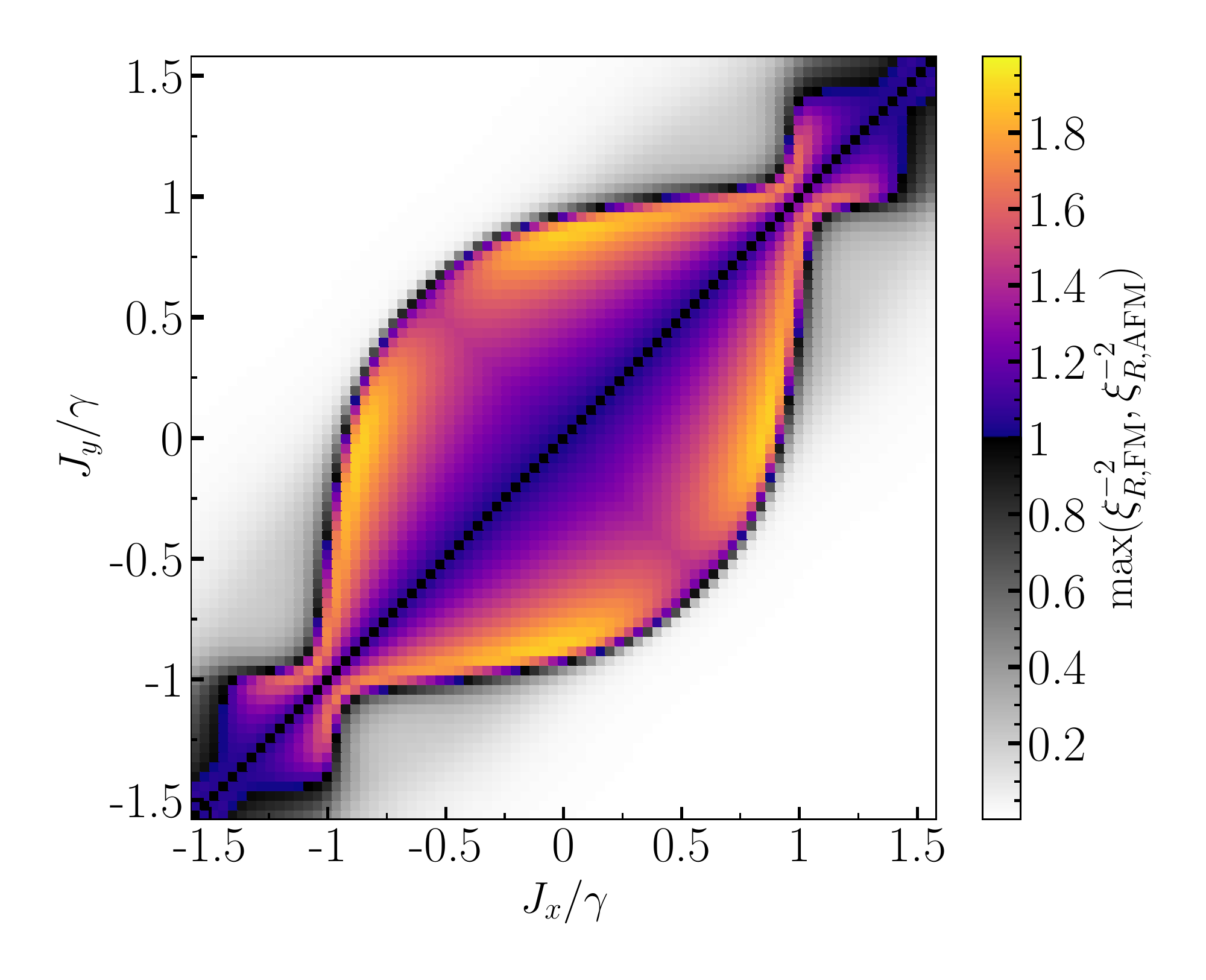}
    \caption{Maximum of the inverse of the spin squeezing parameter $\xi_R^{-2}$ for the collective spin with uniform (FM)  or staggered (AF) collective spin in the $xy$ plane. All simulation parameters as in Fig.~\ref{fig:m2phiN7}.}
    \label{fig:invsqueezN7}
\end{figure}

\begin{figure}
    \centering
    \includegraphics[width=0.5\textwidth]{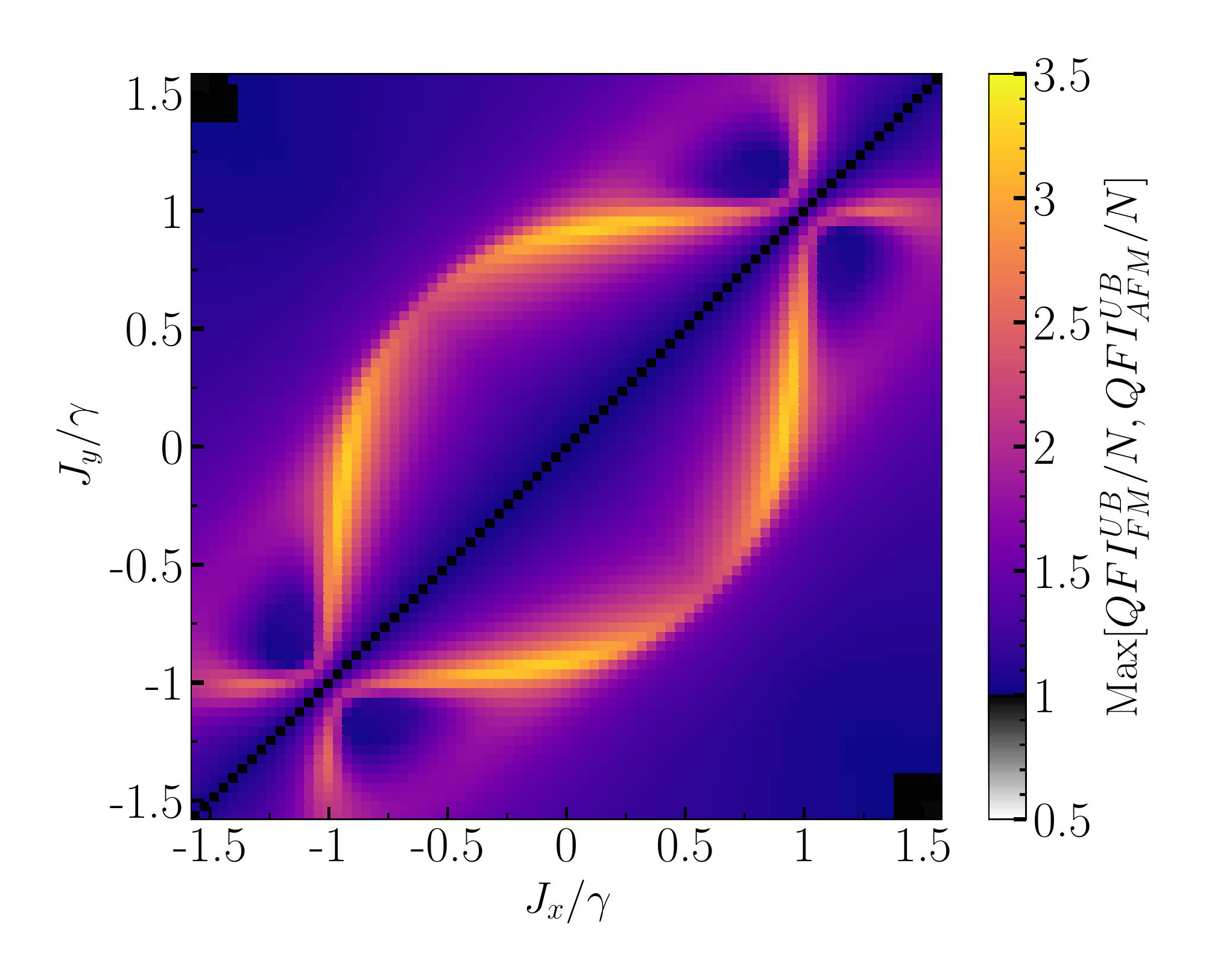}
    \caption{Maximum upper bound to the QFI of collective spin components, $4{\rm max}[F_q(\hat J^\phi), F_q(\hat J_{st}^\phi)]/N$. All simulation parameters as in Fig.~\ref{fig:m2phiN7}.}
    \label{fig:upperboundN7}
\end{figure}

\subsection{Phase diagram: total and quantum correlations} 
\label{sec:phd}

We conclude this section with an overview of total and quantum correlations across the phase diagram of the system at fixed $J_z/\gamma = 1$, and for variable $J_x/\gamma$ and  $J_y/\gamma$. First of all, let us remark that the phase diagram is symmetric under the exchange $J_x \leftrightarrow J_y$, namely it is mirror-symmetric around the $J_x=J_y$ axis. Moreover, there is a mirror symmetry when the signs of both $J_x,J_y$ are switched, and concomitantly ferromagnetic phases in the $xy$ plance are mapped to  antiferromagnetic ones \cite{LeePRL13}. This latter symmetry reflects the fact that the canonical transformation $\sigma_i^{x(y)} \to (-1)^i \sigma_i^{x(y)}$, corresponding to a $\pi$-rotation of one of the two sublattices of the square lattice, leaves the dissipation term unchanged in the GSKL equation; and it therefore establishes a correspondence between the steady states of the system with   couplings $J_x, J_y$ and that of the system with couplings $-J_x, -J_y$.

\subsubsection{Structure factor} 

Fig.~\ref{fig:m2phiN7} shows the evolution of total correlations throughout the phase diagram, as captured by the maximum of the structure factors at the optimal angle $\phi$, 
namely the maximum between $S_{\rm SS}^{\phi\phi}({\bm k}=0)$ (characterizing the ferromagnetic phase) and $S_{\rm SS}^{\phi\phi}({\bm k}=(\pi,\pi))$ (characterizing the antiferromagnetic phase).  The results shown in Fig.~\ref{fig:m2phiN7} have been obtained on a $6\times 6$ lattice: this system size is smaller than the ones used to obtain the results presented above, yet sufficient to capture the overall shape of the phase diagram. We can clearly observe two ferromagnetic and two antiferromagnetic islands, surrounded by paramagnetic regions. The transitions from paramagnetic behavior to ferro/antiferromagnetic order appear rather sharp in the vicinity of the symmetry axis $J_x=J_y$, but much smoother away from it,  revealing that finite-size effects are more pronounced in those ranges of parameters; as a consequence one may erroneously deduce from the study of a finite system that the (anti)ferromagnetic behavior persists for much larger values of $|J_y|$ or $|J_x|$, or that the paramagnetic phase is in fact not re-appearing at all when moving far away from the symmetry axis, as predicted at the mean-field level \cite{LeePRL13}. Yet the existence of a true (anti)ferromagnetic-paramagnetic transition at large $J_y$ (or $J_x$) was firmly established by our results of Sec.~\ref{sec:phasetrans_expon}. 

The FM and AFM islands are connected by an arc-shaped line of maxima in the structure factor. In fact, a cut through one of these arcs already appeared in Fig. \ref{fig:ss_big} around $J_y=0$. The finite size scaling in Fig. \ref{fig:ss_big} (b) showed that the structure factor tends to zero in the thermodynamic limit, and that the origin of the local maximum of the structure factor is an enhancement of quantum fluctuations. This enhancement corresponds to a crossover between a strongly polarized paramagnetic phase ($m^z \gtrsim -1$) and a much more weakly polarized paramagnetic phase ($|m^z| \ll 1$).

 Our analysis did not consider any spin-density-wave phase, which is instead predicted by the mean-field analysis \cite{LeePRL13} in a small region of the diagram at high and slightly unequal $J_x,J_y$. The reason is that its appearence in a finite system is restricted by commensurability of the period with the lattice constant and system length.

\subsubsection{Spin squeezing, upper bound to the quantum Fisher information}

As for the behavior of quantum correlations across the phase diagram,  Fig.~\ref{fig:invsqueezN7} shows the evolution of the inverse squeezing parameter ${\rm max}(\xi_{R,\rm FM}^{-2}, \xi_{R,\rm AF}^{-2})$, where $\xi_{R,\rm FM}^{2}$ is the squeezing parameter defined in Eq.~\eqref{eq:xi2R}, while $\xi_{R,\rm AF}^{2}$ corresponding to the squeezing parameter defined with the variance of the staggered magnetization 
\begin{equation}
\xi^2_{R,\rm AF} = \frac{N~ {\rm Var}(J^{\phi_\perp}_{\rm st})}{\langle J^z \rangle^2}
\end{equation}
where we introduced the staggered total spin $\hat J^{\phi_\perp}_{\rm st} = \sum_i (-1)^i \hat \sigma_i^{\phi_\perp}$ at the angle $\phi_\perp$ perpendicular to the optimal one.  
On the other hand, Fig.~\ref{fig:upperboundN7} shows the evolution of the unravelling-dependent upper bound $4 {\rm max}[F_q(\hat J^{\phi}),F_q(\hat J^{\phi}_{\rm st})]/N$, where we introduced the staggered total spin $\hat J^{\phi}_{\rm st}$ at the optimal angle $\phi$. 
The maximization procedure used for both figures allows us to correctly capture the amount of quantum correlations in the ferromagnetic regime as well as antiferromagnetic one of the phase diagram.

We observe that squeezing is a characteristic of the whole low-entropy paramagnetic regime comprised between the $J_x=J_y$ symmetry axis and the arc-shaped lines of maxima of the structure factor. In particular it becomes very pronounced
along the boundaries of the (A)FM islands which are closest to the  $J_x=J_y$ symmetry axis, and especially so for $J_y/\gamma \lesssim 1$ and $J_x/\gamma \ll 1$, and symmetrically for $J_x/\gamma \lesssim 1$ and $J_y/\gamma \ll 1$; and all along the arcs connecting the FM and AFM islands.

The same regions are also highlighted as being the ones potentially hosting the strongest quantum correlations when looking at the behavior of the upper bound $4 F_q(\hat J^{\phi})$ (somewhat surprising given that this phase is described reasonably well with mean-field theory); with the possibility that the re-entrant paramagnetic phase at large distance from the symmetry axis be also quantum correlated (albeit not squeezed). On the other hand, both $\xi_R^{-2}$ and $4 F_q(\hat J^{\phi})$ single out the ferromagnetic and antiferromagnetic phases as being the ones hosting the weakest quantum correlations in their respective  parameter ranges.

\section{Conclusions \label{sec:concl}}

In this work, we have introduced a new technique for the theoretical study of driven-dissipative many-body spin systems. Our method is based on the combination of the quantum trajectory approach to dissipative evolutions; and of a description of the state along each trajectory based on one- and two-point spin-spin quantum correlation functions only, following a truncation scheme of the cumulant hierarchy for the pure states along each trajectory. Our approach is able to account for both classical and quantum fluctuations at all length scales; and it uniquely makes a simplifying assumption on the statistics of quantum fluctuations along each trajectory. Such an assumption is crucial to limit the computational  cost of our approach to a ${\cal O}(N^2)$ polynomial scaling with the number $N$ of quantum spins (for short-range interactions), making large system sizes ($N\gtrsim 200$) accessible. 

We have applied our method to the dissipative two-dimensional XYZ model, a paradigmatic nonequilibrium system that shows a rich and debated phase diagram. In particular, we find that the re-entrant phase transition from the ferromagnetic to the paramagnetic state upon increasing the coupling of one of the spin components is a true phase transition rather than a crossover. Finite-size scaling of the order-parameter fluctuations conclusively shows that the critical behavior near both transitions belongs to the classical 2D Ising universality class. The classical nature of criticality at the dissipative phase transitions of the model is further supported by our analysis of classical vs. quantum correlations associated with the trajectory unraveling. We find that quantum correlations are always short-ranged, even close to the phase transition, so that the critical behavior is systematically dominated by the classical fluctuations between the different trajectories. Nonetheless, accounting for short-ranged quantum correlations appears to have crucial repercussions on the ensuing critical behavior: indeed neglecting quantum correlations altogether (as in the Gutzwiller-state trajectories) leads to critical behavior incompatible with the 2D Ising one, and in fact rather difficult to analyze in light of known universality classes. The latter holds true even if short-range quantum correlations restricted to 'clusters' are considered.  
Moreover quantum correlations, albeit short-ranged, are still associated with certifiable entanglement related to spin squeezing. This form of entanglement is \emph{enhanced} at the paramagnetic-ferromagnetic transition, showing that the competition between the coherent Hamiltonian dynamics and the incoherent coupling to a bath can in fact induce quantum entanglement in the steady state when tuned in the vicinity of a dissipative critical point. Surprisingly, we also predict significant entanglement in the paramagnetic phase, specifically at the crossover region that replaces the mean-field para-to-(anti)ferromagnetic phase transitions. We note that we have mapped the phase diagram for fixed $J_z/\gamma$, but changing this parameter is expected to lead mostly to a shift of the position of the phases, while leaving the overall features of the phase diagram unchanged. A possible exception could be the limit $J_z/\gamma \rightarrow 0$, where a staggered XY-phase is predicted to appear by mean-field theory \cite{LeePRL13}; yet, a more extensive study of the latter phase \cite{McKeever20} indicates that it may be unstable to fluctuations beyond the mean field approximation. 
Our simulations did not include pure dephasing jumps (Lindblad operators of the form $\hat{\sigma}_z$) either, which might also be an additional effect of the environment \cite{LeePRL13}. However, since we already found that the critical behavior is dominated by classical fluctuations, we do not expect the inclusion of dephasing to change the picture.

In view of the success of our method for the dissipative XYZ model, we expect that it will yield new insights in a variety of dissipative spin systems, thanks to its ability to combine the inclusion of quantum effects at all length scales with the ability to study relatively large systems.

The assumption of the truncation of the correlation hierarchy to two-point correlations may be justified a posteriori by the effect of the environment, preventing quantum correlations from spreading significantly across the system, and from moving to progressively higher orders. Exceptions may exist to this picture, requiring the inclusion of higher-order correlations, such as the study of topological order in dissipative systems, which are known to relate to higher order irreducible correlations \cite{qimqm_2019} 
Extending the method to include $k$-point correlation functions leads to a computational cost scaling as ${\cal O}(N^k)$, which is still manageable, although the sizes that are practically accessible will be necessarily reduced.

The approach that we described is very flexible. In our present work we have focused on the steady state of the dissipative dynamics, but our approach gives as well the possibility of tracking the whole evolution of the system, starting from any initial state which is compatible with the truncation scheme of correlations. It does not rely on assumptions on lattice geometry such as locality, sparsity or symmetries. Application to high-dimensional setups such as arbitrary graphs \cite{Tindall2022}
would be straightforward.
We further note that our method is compatible with a bosonic \cite{VerstraelenAS18,VerstraelenPRR20} or fermionic Ansatz for the trajectory states, opening the way to the study of composite systems comprising different constituents. 
Finally, even in closed systems, the addition of fictitious dissipation has proven to be useful to obtain quantitatively meaningful results \cite{Wouters2020,Somoza2019,Fernandes22}, and our method could be used in that context as well.

In the current stage of development of quantum technologies and the study of driven-dissipative physics accessible to experiments \cite{CarusottoNat2020,chang2014quantum,Leibfried03,Kavokin}, there are a few important tasks: for example
to assess the impact of decoherence and dissipation in realistic quantum simulation / computing setups and to envision novel quantum states stabilized away from equilibrium by the competition between and engineered unitary dynamics and the coupling to an engineered bath. We believe that the approach outlined in this work paves the way towards a systematic investigation of many-body phenomena in open quantum systems, and as such will contribute to the development of quantum technologies with open systems.

\acknowledgements  Discussions with  Rapha\"el Menu, Fabrizio Minganti and Lennart Fernandes; as well as comments on the manuscript and support from Timothy C.H. Liew are gratefully acknowledged. This work was supported by UAntwerpen/DOCPRO/34878. W.V. gratefully acknowledges support from the Singaporean Ministry of Education Tier 2 grant MOE2019-T2-1-004. D. H and T. R. gratefully acknowledge the support of ANR ('EELS' project) and QuantERA ('MAQS' project). Part of the computational resources and services used in this work were provided by the VSC (Flemish Supercomputer Center), funded by the Research Foundation - Flanders (FWO) and the Flemish Government department EWI.

\appendix 
\FloatBarrier
\begin{widetext}

\section{ Evolution equations for the $k=2$ truncation scheme} \label{sec:hetGauss}
 In this section we detail the equations for the evolution of the single-spin and two-spin correlators stemming from the at the basis of the $k=2$ truncation scheme. 
The equations for the single-spin expectation values read:
\begin{align*}
    \dd{\es{s}{x}}= &\left(-\frac{\gamma}{2}\es{s}{x}+2J_y \sum_{s'} \ess{s'}{y}{s}{z} -2J_z \sum_{s'} \ess{s'}{z}{s}{y}\right) \dd{t} \\&+\sqrt{\frac{\gamma}{2}}\left(1+\es{s}{z}-\es{s}{x}^2\right) \dd{W}^x_s+\sqrt{\frac{\gamma}{2}}\es{s}{x}\es{s}{y} \dd{W}^y_s\\
    &+\sqrt{\frac{\gamma}{2}}\sum_{j\neq s}\eff{s}{x}{j}{x}\dd{W}^x_j-\sqrt{\frac{\gamma}{2}}\sum_{j\neq s}\eff{s}{x}{j}{y}\dd{W}^y_j \\
    \dd{\es{s}{y}}=&\left(-\frac{\gamma}{2}\es{s}{y}+2J_z \sum_{s'} \ess{s'}{z}{m}{x} -2J_x \sum_{s'} \ess{s'}{x}{s}{z}\right) \dd{t} \\&-\sqrt{\frac{\gamma}{2}}\es{s}{x}\es{s}{y} \dd{W}^x_s-\sqrt{\frac{\gamma}{2}}\left(1+\es{s}{z}-\es{s}{y}^2\right) \dd{W}^y_s\\
        &+\sqrt{\frac{\gamma}{2}}\sum_{j\neq s}\eff{s}{y}{j}{x}\dd{W}^x_j-\sqrt{\frac{\gamma}{2}}\sum_{j\neq s}\eff{s}{y}{j}{y}\dd{W}^y_j \\
    \dd{\es{s}{z}}=&\left(-\gamma (\es{s}{z}+1)+2J_x \sum_{s'} \ess{s'}{x}{s}{y} -2J_y \sum_{s'} \ess{s'}{y}{s}{x} \right) \dd{t}\\
    &-\sqrt{\frac{\gamma}{2}}\es{s}{x}(1+\es{s}{z})\dd{W}^x_s+\sqrt{\frac{\gamma}{2}}\es{s}{y}(1+\es{s}{z})\dd{W}^y_s\\
        &+\sqrt{\frac{\gamma}{2}}\sum_{j\neq s}\eff{s}{z}{j}{x}\dd{W}^x_j-\sqrt{\frac{\gamma}{2}}\sum_{j\neq s}\eff{s}{z}{j}{y}\dd{W}^y_j \\
\end{align*}
where $\delop_s^\alpha=\sop_s^\alpha-\es{s}{\alpha}$ and consequently $\eff{s}{\alpha}{m}{\beta}=\ess{s}{\alpha}{m}{\beta}-\es{s}{\alpha}\es{m}{\beta}$  indicates the two-site covariance.

Eq. \eqref{eq:hettraj-exp} leads to several contributions to the evolution of the covariances: two terms from the deterministic parts; and an Ito term and two noise terms from the stochastic part:
\begin{equation}
    \dd \ev{ \delop_s^\alpha \delop_m^\beta } 
    =\ev{ \left( \dd_D \delop_s^\alpha \right) \; \delop_m^\beta } 
    + \ev{  \delop_s^\alpha \; \left( \dd_D \delop_m^\beta \right)}  
    + d_I \ev{  \delop_s^\alpha \; \delop_m^\beta} \textbf{}
    + \ev{ \left( \dd_S \delop_s^\alpha \right) \; \delop_m^\beta  } 
    + \ev{    \delop_s^\alpha \; \left( \dd_S \delop_m^\beta  \right)}.
\end{equation}
For the deterministic contributions, one can simply substitute
\begin{align}\label{eq:detcovterms}
\dd_D\delop_s^x &=\left(-\frac{\gamma}{2} \delop_s^x + 2J_y\sum_{s'} \sop^z_s \sop^y_{s'} -2J_z\sum_{s'} \sop^y_s \sop^z_{s'} \right)\dd{t} \nonumber\\
\dd_D \delop_s^y &=\left(-\frac{\gamma}{2} \delop_s^y +2J_z\sum_{s'} \sop^x_s \sop^z_{s'} -2J_x\sum_{s'} \sop^z_s \sop^x_{s'} \right)\dd{t} \nonumber\\
\dd_D \delop_s^z &=\left(-\gamma\delop_s^z+2J_x\sum_{s'} \sop^y_s \sop^x_{s'} - 2J_y\sum_{s'} \sop^x_s \sop^y_{s'} \right)\dd{t} \nonumber\\
\end{align}

For the stochastic terms, one has
\begin{align}\label{eq:stochcovterms}
{ \ev{  \left( \dd_S\delop_s^x \right) \; \cdot}}&=
\sqrt{\frac{\gamma}{2}}\left(-2\ev{\sop^x_s}\ev{\delop_s^x\cdot}+\ev{\delop_s^z\cdot}\right)\dd{W}^x_s+\sqrt{\frac{\gamma}{2}}\left(\ev{\sop^y_s}\ev{\delop_s^x\cdot}+\ev{\sop^x_s}\ev{\delop_s^y\cdot}\right)\dd{W}^y_s \nonumber\\
{\ev{ \left(  \dd_S \delop_s^y \right) \; \cdot}}&=
-\sqrt{\frac{\gamma}{2}}\left(\ev{\sop^x_s}\ev{\delop_s^y\cdot}+\ev{\sop_s^y}\ev{\delop_s^x\cdot}\right)\dd{W}^x_s+\sqrt{\frac{\gamma}{2}}\left(2\ev{\sop^y_s}\ev{\delop_s^y\cdot}-\ev{\delop_s^z\cdot}\right)\dd{W}^y_s \nonumber\\
{\ev{ \left( \dd_S\delop_s^z \right) \; \cdot}}&=
-\sqrt{\frac{\gamma}{2}}\left(\ev{\delop_s^x\cdot}+\ev{\sop_s^x}\ev{\delop_s^z\cdot}+\ev{\sop_s^z}\ev{\delop_s^x\cdot}\right)\dd{W}^x_s+\sqrt{\frac{\gamma}{2}}\left(\ev{\delop_s^y\cdot}+\ev{\sop_s^y}\ev{\delop_s^z\cdot}+\ev{\sop_s^z}\ev{\delop_s^y\cdot}\right)\dd{W}^y_s  
\nonumber\\
{\ev{\cdot \; \left( \dd_S \delop_m^x \right) }}&= 
\sqrt{\frac{\gamma}{2}}\left(-2\ev{\sop^x_m}\ev{\cdot\delop_m^x}+\ev{\cdot\delop_m^z}\right)\dd{W}^x_m+\sqrt{\frac{\gamma}{2}}\left(\ev{\sop^y_m}\ev{\cdot\delop_m^x}+\ev{\sop^x_m}\ev{\cdot\delop_m^y}\right)\dd{W}^y_m \nonumber\\
{\ev{\cdot \; \left( \dd_S \delop_m^y  \right) }}&=
-\sqrt{\frac{\gamma}{2}}\left(\ev{\sop^x_m}\ev{\cdot\delop_m^y}+\ev{\sop_m^y}\ev{\cdot\delop_m^x}\right)\dd{W}^x_m+\sqrt{\frac{\gamma}{2}}\left(2\ev{\sop^y_m}\ev{\cdot\delop_m^y}-\ev{\cdot\delop_m^z}\right)\dd{W}^y_m \nonumber\\
{\ev{\cdot \; \left( \dd_S\delop_m^z \right)}}&=
-\sqrt{\frac{\gamma}{2}}\left(\ev{\cdot\delop_m^x}+\ev{\sop_m^x}\ev{\cdot\delop_m^z}+\ev{\sop_m^z}\ev{\cdot\delop_m^x}\right)\dd{W}^x_m+\sqrt{\frac{\gamma}{2}}\left(\ev{\cdot\delop_m^y}+\ev{\sop_m^y}\ev{\cdot\delop_m^z}+\ev{\sop_m^z}\ev{\cdot\delop_m^y}\right)\dd{W}^y_m,
\end{align}
where the `$\cdot$' can be replaced with any operator $\delop_m^\alpha$.
Finally, the Ito terms give contributions such as
\begin{align}\label{eq:Itoterms}
\dd_I{\eff{s}{x}{m}{x}}&=-\frac{\gamma}{2}\left[\left(1+\es{s}{z}-\es{s}{x}^2\right)\eff{m}{x}{s}{x}+\es{s}{x}\es{s}{y}\eff{m}{x}{s}{y}+\left(s\leftrightarrow m\right)\right]\dd{t}   \nonumber\\
&-\frac{\gamma}{2}\sum_{j\neq s,m}\left[\eff{s}{x}{j}{x}\eff{j}{x}{m}{x}+\eff{s}{x}{j}{y}\eff{j}{y}{m}{x}\right]\dd{t}.
\end{align}

Equations \eqref{eq:detcovterms} contain the  three-spin terms of the form $\ev{\sop_s^{\alpha}\sop_{s'}^\beta \delop_m^\gamma}$. We can reduce them to functions of single- and two-spin terms by assuming the vanishing of the third-order cumulant; in doing so, we distinguish the cases whether $s'$ index is equal to $m$ or not.
If $s'=m$,  $\sop_m^\beta\sop_m^\gamma=\sop_m^{\beta\gamma}$ is another single spin operator and
\begin{equation}
\ev{\sop_s^{\alpha}\sop_{m}^\beta \delop_m^\gamma}=\eff{s}{\alpha}{m}{\beta\gamma}+\es{s}{\alpha}\es{m}{\beta\gamma}-\es{m}{\gamma}\ess{s}{\alpha}{m}{\beta}.    
\end{equation}
otherwise,
\begin{equation}
\ev{\sop_s^{\alpha}\sop_{s'}^\beta \delop_m^\gamma}=\es{s}{\alpha}\eff{s'}{\beta}{m}{\gamma}+\es{s'}{\beta}\eff{s}{\alpha}{m}{\gamma}.    
\end{equation}

Similarly,
\begin{equation}
\ev{\delop_s^{\alpha}\sop_{s}^\beta \sop_m^\gamma}=\eff{s}{\alpha\beta}{m}{\gamma}+\es{s}{\alpha\beta}\es{m}{\gamma}-\es{s}{\alpha}\ess{s}{\beta}{m}{\gamma}    
\end{equation}
and for $s\neq m'$
\begin{equation}
\ev{\delop_s^{\alpha}\sop_{m'}^\beta \sop_m^\gamma}=\es{m'}{\beta}\eff{s}{\alpha}{m}{\gamma}+\es{m}{\gamma}\eff{s}{\alpha}{m'}{\beta}.    
\end{equation}


\section{Finite measurement efficiencies\label{ap:measeff}}
We have found that the numerical simulation of the quantum trajectory $k = 2$ equations are subject to numerical instabilities in various parameter regimes. More specifically, the expectation values tend to diverge at very short time scales, leading to unphysical results and ultimately numerical ``Not a Number'' (NaN) results. These numerical instabilities are caused by the noise terms coupled to the second-order cumulants (Eqs.\eqref{eq:stochcovterms}), or second-order noise terms in short. 
and are very hard to control or suppress. However, not including the second-order noise terms allows one to simulate the equations and yield physical results with little to no numerical instabilities. 
We now show, through numerical example, that these second-order noise terms are not important, and can be neglected from the equations, by using the concept of measuring efficiency. This method will allow us to profit from the numerical stability of the standard master equation approach, in combination with the increase in the method's complexity due to the trajectory approach. We first recall the concept and will then apply it to the $XYZ$ model studied in the main text, of which the results will be shown in Fig.~\ref{fig:App-m2N7}.

In the quantum trajectory formalism one assumes the existence of perfect detectors continuously monitoring the system of interest. This leads to Eq.~\eqref{eq:hettraj}. However, one can equivalently derive equations for the more general case where the detector has a finite efficiency $0\leq \eta\leq1$ \cite{Wiseman_BOOK_Quantum}.

 Indeed, one can re-write Eq.~\eqref{eq:GKSL} as
\begin{equation}\label{eq:app:fme}
    \partial_t \Rho =-i\comm{\Ham}{\Rho}+\frac{1}{2}\sum_j(1-\eta)\left(2\Gamop_j\Rho\Gamop_j^\dagger-\Gamop_j^\dagger\Gamop_j\Rho-\Rho\Gamop_j^\dagger\Gamop_j\right)+\frac{1}{2}\sum_j\eta\left(2\Gamop_j\Rho\Gamop_j^\dagger-\Gamop_j^\dagger\Gamop_j\Rho-\Rho\Gamop_j^\dagger\Gamop_j\right),
\end{equation}
where we have re-written the original dissipator as two individual dissipators with Lindblad operators $\sqrt{1-\eta}\hat{\Gamma}_j$ and $\sqrt{\eta}\hat{\Gamma}_j$. 
We can imagine now that only the part proportional to $\eta$ contributes to the noise term in the stochastic Schr\"odinger's equation emerging in the quantum-state-diffusion approach, Eq.~\eqref{eq:hettraj}. 
Physically, this process corresponds to a situation in which the information (i.e. photon current) leaking out of the system is collected with a finite efficiency $\eta\leq 1$. One can then straightforwardly show that this yields the following equation for the expectation value of an operator $\hat{O}$ 
 \begin{align}
\dd{\ev{\Oop}} & =i \ev{\comm{\Ham}{\Oop}}\dd{t}
-\frac{1}{2}\sum_j\left(\ev{\Gamop_j^\dagger\comm{\Gamop_j}{\Oop}}-\ev{\comm{\Gamop_j^{\dagger}}{\Oop}\Gamop_j}\right)\dd{t} \\ 
&+\sqrt{\eta}\sum_j \left(\ev{\Gamop_j^\dagger (\hat O -\langle \hat O \rangle)}\dZ{j}
+ \ev{(\hat O -\langle \hat O \rangle) \Gamop_j}\dZ{j}^*\right). \nonumber
\end{align}
Including a finite efficiency for the continuous monitoring process thus only leads to a factor $\sqrt{\eta}$ in the noise part of the trajectory formalism. A measuring efficiency $\eta = 1$ brings us back to the original quantum trajectory equations; while for $\eta = 0$ one retrieves the master equation approach to correlation hierarchies (Eq.~\eqref{eq:masterhierarchy}). For the latter, no noise terms are present (on any cumulant) and the earlier mentioned numerical instabilities are absent In practice, the imperfect measuring efficiency $\eta$ makes the simulation of the equations numerically more stable. As a result, it allows one to numerically solve the equations using the $k = 2$ quantum trajectory formalism until the numerical instability becomes unmanageable. Note that such a formalism is physical by itself \cite{Wiseman_BOOK_Quantum}, and naturally compatible with our approach.

Hence, we will use this finite measuring efficiency formalism to solve the $k = 2$ quantum trajectory equations for various values of $\eta$ and show that their results are identical to the $k = 2$ quantum trajectories \emph{without} second-order noise terms (and of course with the same respective $\eta$ i.e. noise coefficients to the first order cumulants). Such an analysis allows us to extrapolate the validity of the $k = 2$ quantum trajectory approach without second-order noise terms at full measuring efficiency ($\eta = 1$), and thus in the regime where one profits most of the trajectory approach with respect to the master equation approach ($\eta = 0$). 

\begin{figure}
    \centering
    \includegraphics[width=\textwidth]{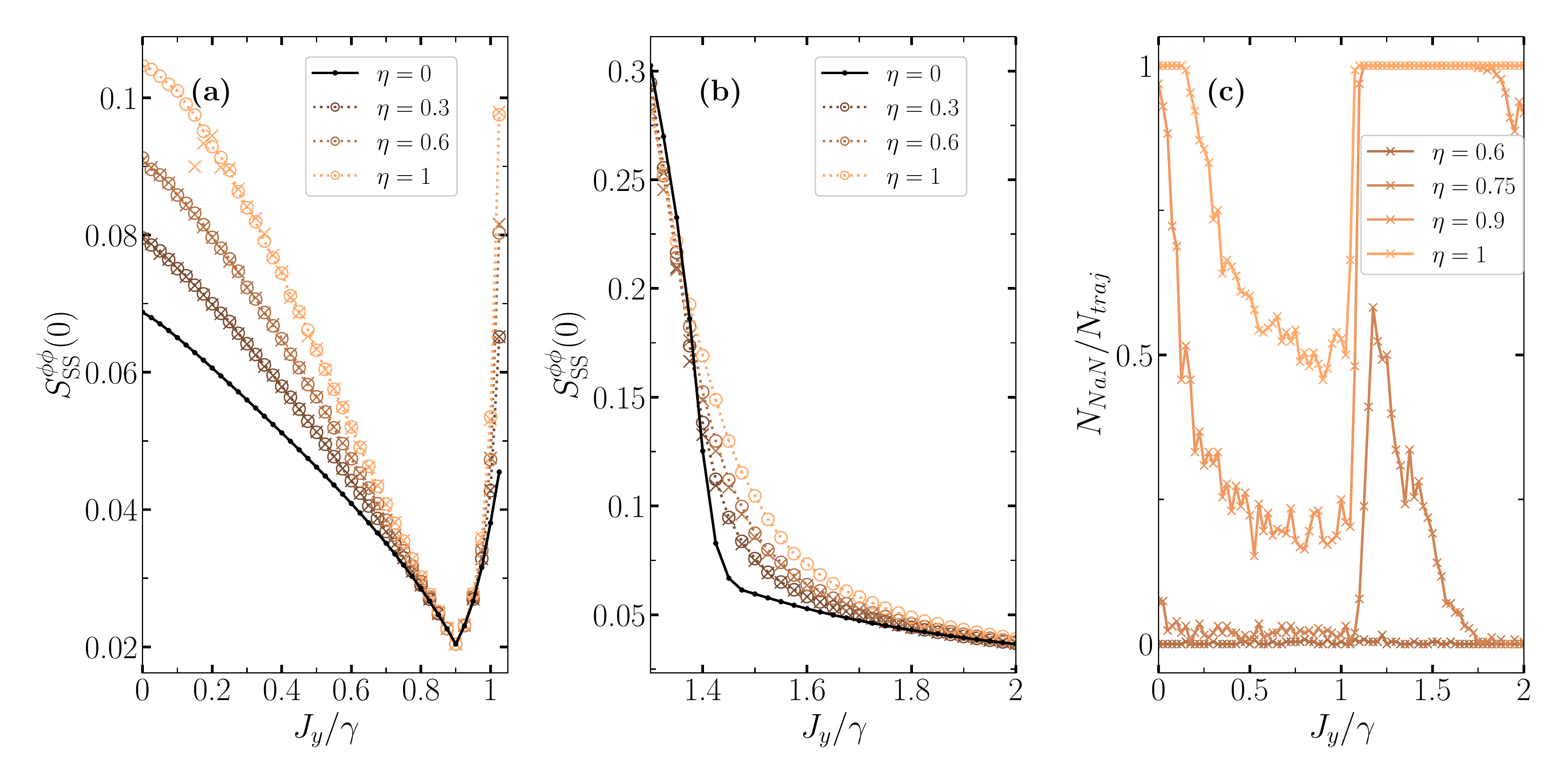}
    \caption{
    \textbf{(a)} - \textbf{(b)} Steady-state structure factor $S^{\phi\phi}_{\rm SS}(0)$ on a $7\times7$ lattice for various measuring efficiencies $\eta$, with second-order noise contributions (cross markers) and without second-order noise contributions (circle markers with dotted lines), for two regions of the phase diagram. 
    \textbf{(c)} number of diverging, i.e. NaN-trajectories, $N_{\rm NaN}$ with respect to the total number of trajectories $N_{\rm traj}$. Only results with second-order noise terms (full lines with crosses) are shown as results without second-order noise terms are all located at (approximately) zero.
    The number of trajectories for the results in all panels is $N_{\rm traj} = 256$, and time averages are taken  over the time interval $t\gamma \in \left[75; 150\right]$.}
    \label{fig:App-m2N7}
\end{figure}

We will now revisit the results for the dissipative XYZ model, by focusing on the steady-state structure factor $S_{\rm SS}^{\phi\phi}(0)$ for the $\phi$ spin components displaying the strongest correlations (see Sec.~\ref{sec:structfact}). In Fig.~\ref{fig:App-m2N7} (a) and (b) we show the the $k = 2$ quantum trajectory results with and without second-order noise contributions (Eqs.\eqref{eq:stochcovterms}) for various values of $\eta$ of a $7\times7$ lattice. 
 We generally observe a very good agreement between the results excluding the second-noise terms and those including them (when available) for various values of $\eta$, showing that those noise contributions are in fact negligible whenever they do not lead to numerical instabilities.

Note that for higher values of $\eta$ occasional results deviate for values of $J_y/\gamma \approx 0.1$, see Fig.~\ref{fig:App-m2N7} (a), which is due to the increasing number of diverging trajectories. We call this number $N_{\rm NaN}$ and show their rate with respect to the total number of trajectories $N_{\rm traj}$ in Fig.~\ref{fig:App-m2N7} (c). Every time a trajectory diverges, we discard the entire trajectory and do not use it to gather statistics on the system. We note that such omission of divergent trajectories is  mathematically justified under quite general conditions \cite{milstein_tretyakov}.
Due to the very low number of non-diverging trajectories at high measuring efficiencies, the gathered statistics will evidently be low. Hence some deviations from the (stable) results where the second-order noise term has been omitted appear in the finite numerics. For example, for the highest values of $\eta$, the number of diverging trajectories becomes equal to the number of simulated trajectories for the highest (and lowest) values of $J_y/\gamma$ shown on Fig.~\ref{fig:App-m2N7} (c). This in turn results in a lack of results for the $k=2$ quantum trajectory approach with noise terms, i.e. the original problem.

Nevertheless, as the efficiency $\eta$ is increased, across the parameter regime, one still observes the correspondence between the results with and without noise term. Extrapolating these results in the limit of $
\eta\rightarrow 1$ (i.e. assuming that the second-order noise terms remain negligible at efficiency $\eta=1$  over the entire range of parameters that we explored -- including in parameter regions for which all trajectories become numerically unstable when second-order noise terms are included), we can conclude that the omission of the second-order noise terms is legitimate. All the results presented in the main text for the $k=2$ truncation scheme are therefore solutions to equations with $\eta = 1$  with the omission of second-order noise terms. 

\section{Second and fourth moments of the $xy$ spin components}\label{app:m2m4}
We show results for the (steady-state) second and fourth moments of $\frac{1}{N}\sum_j\hat{\sigma}_j^x$ and $\frac{1}{N}\sum_j\hat{\sigma}_j^y$ in Fig.~\ref{fig:App-m2m4} (a-d). Note that the second moment is identical to the structure factor from Eq.~\eqref{eq:steadystatespinstructurefactor} $S_{SS}^{\alpha\alpha}(\textbf{k}=0) = m_2^\alpha$. The remarkable correspondence with the exact results shown in Fig.~\ref{fig:App-m2m4}~(a) is thus identical to the one discussed in Fig.~\ref{fig:comp_exact}~(a). Nonetheless, the panels (b-d) of  Fig.~\ref{fig:App-m2m4} show that this correspondence to the exact results is not limited to the second moment,  but also persists in the fourth moment for both the $\hat\sigma^x$ and $\hat \sigma^y$ spin components.  Moreover, the correspondence of the second moment $m_2^y$ and the fourth moment $m_4^y$ is even more convincing than that for $m_2^x$ and $m_4^x$. 

The agreement of the fourth moment of the spin components with exact diagonalization shows that the truncation scheme of the cumulant hierarchy to second order does \emph{not} lead to a significant loss in accuracy when looking at higher-order correlators. This observation vindicates the working assumption underlying our approach, namely the fact that cumulants of order higher than $k=2$ are essentially negligible in the steady state.

\begin{figure}
    \centering
    \includegraphics[width=\textwidth]{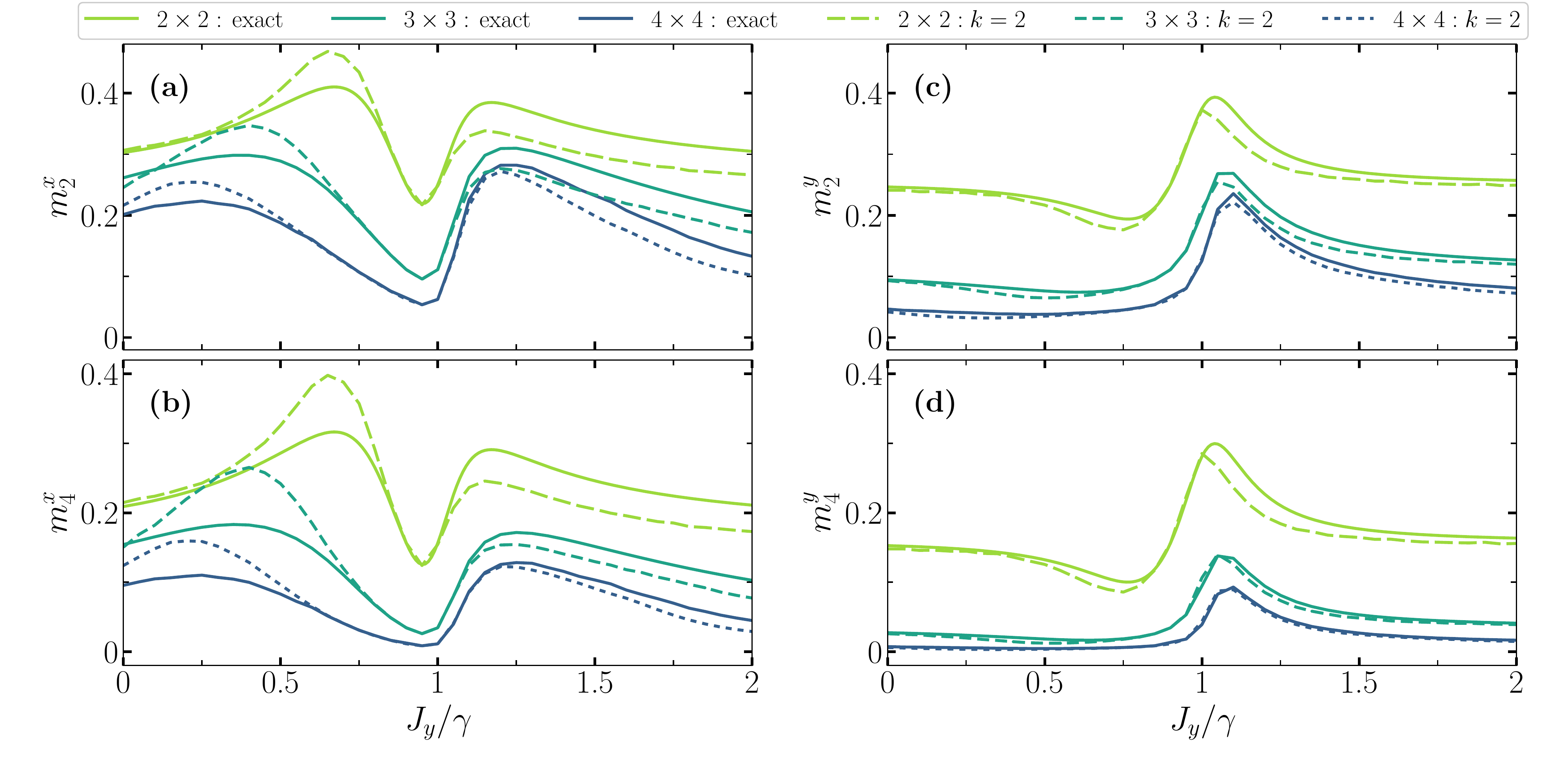}
    \caption{Second and fourth moment for the $\hat\sigma^x$ and $\hat\sigma^y$ spin components obtained with the $k=2$ truncation scheme (dashed lines), and compared with the exact solution (full lines) for three different system sizes. The exact results correspond to a direct solution of the master equation for $N < 4$, and to its stochastic unraveling for $N = 4$. The $k = 2$ results were obtained using $N_{\rm traj} \approx 250$ trajectories, with time-averaging performed over the time interval $t\gamma \in \left[75; 150\right]$.}
    \label{fig:App-m2m4}
\end{figure}

\section{Finite-size scaling of the $k = 1$ results \label{ap:k1}}

In the main text, using the $k=2$ cumulant truncation scheme, we have shown that the phase transitions belong to the universality class of the classical 2D Ising model, and hence the critical fluctuations are of a classical nature. It may thus be tempting to think that the $k=1$ (Gutzwiller) trajectories are already sufficient to describe this behavior, as they are expected to capture classical fluctuations without any a priori assumption or limitation on their spatial structure.

However, as we can see in Fig. \ref{fig:k=1criticality} for the paramagnetic-to-ferromagnetic transition (at $J_x/\gamma = 0.9$, $J_y/\gamma \approx 1$)  the $k=1$ results are incompatible with the 2D Ising universality class. 

{Fig.~\ref{fig:k=1criticality} (a) and (b) show the 
rescaled structure factor $S^{\phi\phi}_{\rm SS}(0) L^{2\beta/\nu}$ using 2D classical Ising exponents exponents, which do not lead to a clear collapse of the curves for different system sizes, even in the vicinity of the putative critical point}. In fact, different combinations of critical exponents can give a better collapse. 
 In particular, a decent collapse is obtained when one takes $\beta = \nu = 1/2$, i.e. the mean field critical exponents, as shown in Fig.~\ref{fig:k=1criticality} (c) and (d). 

For the second transition (from ferromagnetism back to paramagnetism) at even higher $J_y$-values, it was already clear from earlier works that the Gutzwiller trajectory approach \cite{CasteelsPRA18} and its cluster extensions \cite{HuybrechtsPRA19} predict a size-dependent sudden drop in the structure factor instead of a set of smooth curves that find a common crossing upon rescaling, which would be the expected behavior at a continuos phase transition.

\begin{figure}
    \centering
    \includegraphics[width=\textwidth]{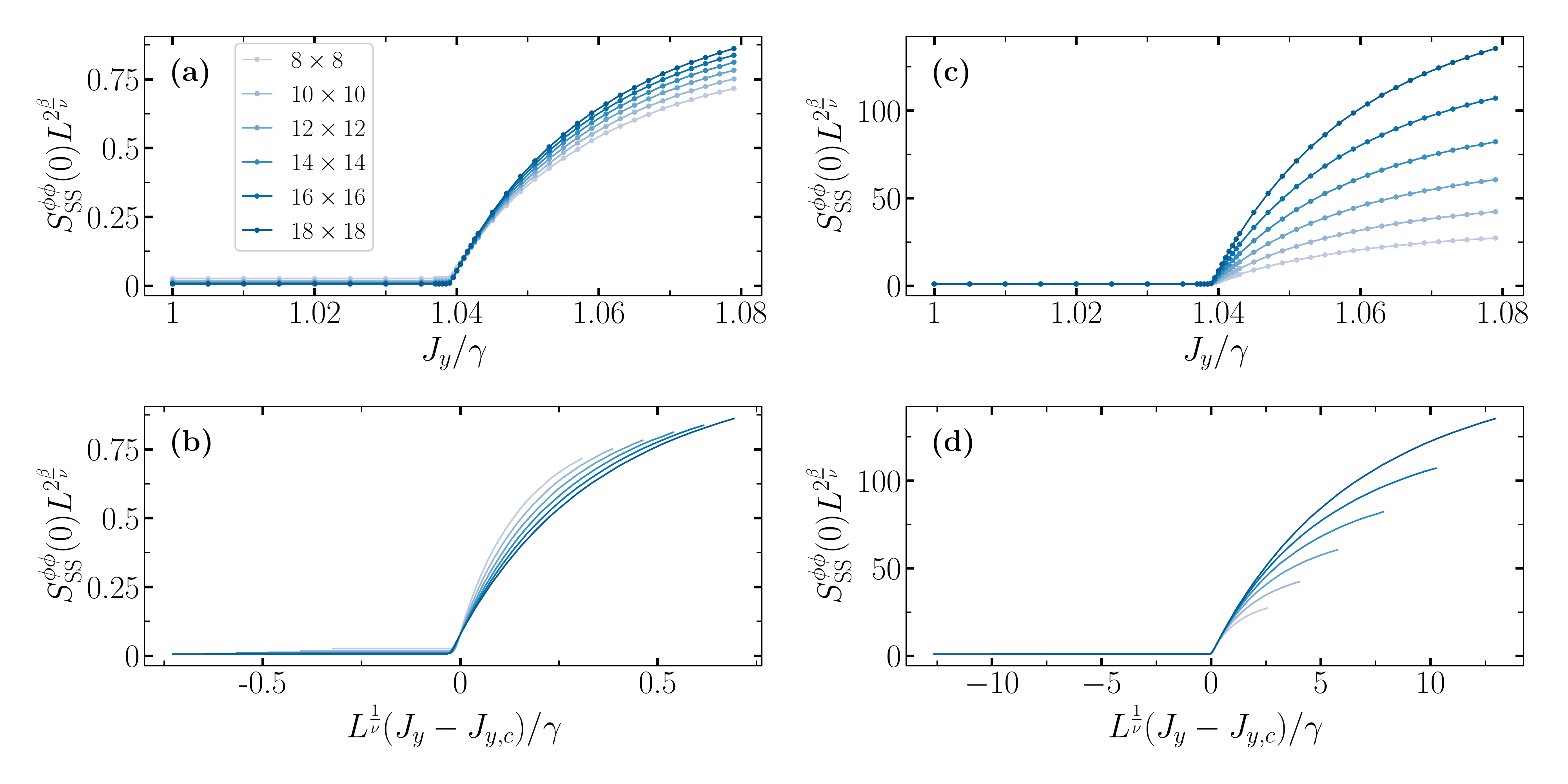}
    \caption{Finite-size scaling analysis of the $k=1$ results for the  structure factor $S_{\rm SS}^{\phi\phi}(0)$: \textbf{(a)} rescaled structure factor $S_{\rm SS}^{\phi\phi}(0) L^{2\beta/\nu}$ using 2D Ising exponents ($\nu = 1$ and $\beta = 1/8$) ; \textbf{(b)} full scaling plot  with the same exponents, using the critical point $J_{y,c} = 1.0405$;    \textbf{(c)} rescaled structure factor $S_{\rm SS}^{\phi\phi}(0) L^{2\beta/\nu}$ using mean-field exponents ($\nu = 
    1/2$ and $\beta = 1/2$); \textbf{(d)} full scaling plot with the same exponents, using $J_{y,c} = 1.039$. All results stem from $N_{\rm traj} \approx 32$ trajectories, and they were time-averaged over the time interval $t\gamma \in \left[700; 1000\right]$.}
    \label{fig:k=1criticality}
\end{figure}

\section{Scaling of the transverse magnetisation derivative at the 2D Ising transition \label{ap:z-mag}}

 In this section we discuss the expected scaling of the derivative of the transverse magnetization with respect to the control parameter of the transition at the thermal phase transition of the 2D Ising model. This discussion serves as a basis for the scaling analysis proposed in Sec.~\ref{sec:phasetrans_expon} for this quantity at the dissipative transition of the 2D XYZ model.   

For the sake of definiteness, we shall use as a reference model the transverse field Ising model, with Hamiltonian 
\begin{equation}
{\hat H} = -J \sum_{\langle ij \rangle} \hat\sigma_i^x \hat\sigma_j^x - \Gamma \sum_i \hat \sigma_i^z~
\end{equation}
defined on the same square lattice as the XYZ model investigated in the main text. Although the above model has a well-known quantum phase transition in the ground state, we shall only focus on its thermal properties, and in particular on the fact that it has a line of thermal 2D Ising transitions at temperatures $T_c(\Gamma)$ which decrease with increasing $\Gamma$, and eventually vanish at the quantum critical point. 

In the vicinity of the critical line, the singular part of the free-energy density is expected to scale as
\begin{equation}
f_s(T,\Gamma) \sim  |T-T_c(\Gamma)|^{2-\alpha}
\end{equation}
which implies that the transverse magnetization $m^z = \langle \sigma_i^z \rangle$ has a singular part going as
\begin{equation}
m^z = - \frac{\partial f_s}{\partial \Gamma} \sim (2-\alpha) \dv{T_c}{\Gamma} ~|T-T_c(\Gamma)|^{1-\alpha}~;
\end{equation}
while the derivative of this magnetization with respect to the control parameter of the transition (namely the temperature $T$) exhibits a singularity going as 
\begin{equation}
\frac{\partial m^z}{\partial T} = - \frac{\partial^2 f_s}{\partial \Gamma \partial T} \sim (2-\alpha)(1-\alpha) \dv{T_c}{\Gamma} ~|T-T_c(\Gamma)|^{-\alpha}
\end{equation}
namely it is has the same singular behavior as the specific heat
\begin{equation}
c_v = -T \frac{\partial ^2 f_s}{\partial T^2} \sim (2-\alpha)(1-\alpha)  |T-T_c(\Gamma)|^{-\alpha}~.
\end{equation}

The 2D Ising universality class has $\alpha=0$, meaning that the scaling dimension of the specific heat, $\alpha/\nu$, is also vanishing. Yet this result still leaves the room for a specific heat diverging at the transition as $\log L$, where $L$ is the linear size of the lattice. Hence we expect that this same scaling property is shared as well with the temperature derivative of the transverse magnetization $\frac{\partial m^z}{\partial T}$. We have verified that this is indeed the case on quantum Monte Carlo data for the thermal transition of the 2D Ising model in a transverse field. We can therefore expect the same scaling behavior to be shared with $\gamma(\dd{m^z}/\dd{J_y})$ at the dissipative transition of the 2D XYZ model if this transition is to comply with the 2D Ising universality class; and indeed the logarithmic divergence of $\frac{\partial m^z}{\partial T}$ is clearly exhibited in Sec.~\ref{sec:phasetrans_expon}  of the main text.

\end{widetext}

\end{document}